\RequirePackage{ifpdf}
\documentclass[12pt]{article}
\usepackage{epsfig, graphicx, bm, amsmath, amsfonts}
\usepackage{epstopdf}
\usepackage{amssymb, xcolor, float}
\usepackage{amsthm}
\usepackage{hyperref}
\usepackage[english]{babel}
\usepackage[T1]{fontenc}
\usepackage[utf8]{inputenc}
\usepackage{authblk}
\usepackage{mathtools}
\usepackage{slashed}
\usepackage{mattens,amssymb}
\usepackage{amsmath,amssymb}
\usepackage{amsfonts}

\definecolor{burgundy}{rgb}{0.5, 0.0, 0.13}

\usepackage{hyperref}
\hypersetup{colorlinks=true,linkcolor=blue,citecolor=red,urlcolor=blue,linktocpage}
\usepackage[capitalize]{cleveref}
\numberwithin{equation}{section}
\allowdisplaybreaks

\usepackage[autostyle]{csquotes}
\usepackage[style=ieee, backend=bibtex, citestyle=numeric-comp, sorting=none]{biblatex} 
\usepackage{graphicx}
\graphicspath{ {./images/} }
\usepackage{subfig}

\usepackage{titlecaps}

\begin{document}

\parindent=12pt
\begin{center}
    {\Large \textbf{Bulk-Boundary Thermodynamics of Charged \\ \vspace{1.5mm} Black Holes in Higher Derivative Theory}}
 \end{center}

\baselineskip=20pt
\bigskip
\centerline{\textbf{Gurmeet Singh Punia}}
\bigskip
\centerline{\it Department of Physics, Indian Institute of Science Education and Research Bhopal}
\centerline{\it Bhopal bypass, Bhopal 462066, India}
\bigskip
\centerline{E-mail: 
\href{mailto:gurmeet17@iiserb.ac.in}{gurmeet17@iiserb.ac.in}}

\vspace*{4.0ex}
\centerline{\today}
\vspace*{4.0ex}

\thispagestyle{empty}

\centerline{\bf Abstract} \bigskip
\noindent 
The connection between bulk and boundary thermodynamics in Einstein-Maxwell theory is well established using AdS/CFT correspondence. In the context of general higher derivative gravity coupled to a U(1) gauge field, we examine the resemblance of the first law of thermodynamics between bulk and boundary, followed by an extended phase space description on both sides. Higher derivative terms related to different powers of the string theory parameter $\alpha'$ emerged from a consistent truncation in the bulk supergravity action.
We demonstrate that one must include the fluctuation of $\alpha'$ in the bulk thermodynamics as a bookkeeping tool to match the bulk first law and Smarr relation with the boundary side. Consequently, the Euler relation and the boundary first law are altered by adding two central charges ($\mathtt{a}$, $\mathtt{c}$). To support our general conclusion, we consider the black hole in Gauss-Bonnet gravity and the general four-derivative theory. Finally, we examine the bulk and boundary aspects of the extended phase space description for higher derivative corrected black holes.
\vfill\eject

\setcounter{page}{1}
\noindent\makebox[\linewidth]{\rule{\textwidth}{1pt}}
\tableofcontents
\noindent\makebox[\linewidth]{\rule{\textwidth}{1pt}}
\section{Introduction and Summary}\label{sec:intro}

The thermodynamics of black holes in Anti-de Sitter (AdS) space has remained fascinating since the publication of Hawking and Page's seminal observations in \cite{Hawking:1982dh}. Then, the Anti-de Sitter/Conformal field theory (AdS/CFT) correspondence, \textit{i.e.} holographic duality enhances our understanding of the AdS black hole's thermodynamics, where observed that the thermodynamic properties of black holes could be reinterpreted as a conformal field theory at finite temperature \cite{Maldacena:1997re, Witten:1998qj}. The thermal properties of Einstein-Maxwell AdS theory \textit{i.e.} Reissner–Nordstr\"{o}m(RN) black hole \cite{Hawking:1995ap} in AdS show an intriguing phase space description, and at the same time, the thermal phase structure of the dual field theory also fascinating via AdS/CFT correspondence \cite{Chamblin:1999tk_charge_BH, Chamblin:1999tk_charge_BH_2}.

The charged black hole thermodynamic phase structure in AdS is enhanced after treating the cosmological constant $\Lambda_0$ as the thermodynamic pressure $P_0 = - \Lambda_0 / 8\pi G_N $ and its conjugate quantity as the thermodynamic volume. Likewise, the phase space is comprehended as \emph{extended phase space}. The inclusion of the variation of the cosmological constant in the first law will complete the analogy of the charged AdS black hole system as the Van der Waals system \cite{Kastor:2009wy, Kubiznak:2012wp,  Altamirano:2013uqa, Altamirano:2013ane, Dutta:2013dca, Johnson:2014yja, Kubiznak:2014zwa, Kubiznak:2016qmn}. This new paradigm is dubbed as \emph{black hole chemistry}\footnote{A higher-dimensional origin of extended black hole thermodynamics \textit{i.e.} black hole chemistry is discussed in \cite{Frassino:2022zaz}} \cite{Kubiznak:2014zwa, Kubiznak:2016qmn}, and as the black hole provides a dual description of the field theory on the boundary in the context of AdS/CFT correspondence. As a consequence, it is envisaged that the thermodynamic variables and laws on both sides would line up. In  \cite{Karch:2015rpa, Visser:2021eqk, Cong:2021fnf, Cong:2021jgb, Dutta:2022wbh} (following the earlier work \cite{Kastor:2010gq}), it demonstrated that the inclusion of the variation of Newton's constant together with the cosmological constant needed as bookkeeping tool if one wants to construct the bulk first law and Smarr relation consistent with boundary thermodynamics. 

Let's describe the current situation regarding the first law and the Smarr relation for charged AdS black holes with two derivative gravity before proceeding any further.  The extended first law of charged AdS black holes in $d+1$ dimension, including the variation of cosmological constant $\Lambda_0$ \cite{Kastor:2009wy} and Newton's constant $G_N$ \cite{Visser:2021eqk, Cong:2021fnf, Cong:2021jgb} is
\begin{eqnarray}\label{eq:bulkfirstlaw}
	\begin{aligned}
		d M_0 = \frac{T_0}{4 G_N} dA_0 + \Phi_0 dQ_0 + \frac{\Theta_0}{8 \pi G_N} d\Lambda_0 - \left(M_0 - \Phi_0 Q_0 \right) \frac{dG_N}{G_N} \,,
	\end{aligned}
\end{eqnarray}
where $M_0$ is the Arnowitt-Deser-Misner mass (ADM mass) of the black hole, $T_0$  is the Hawking temperature, $A_0$ is the area of the event horizon, $Q_0$ is the electric charge, and its conjugate quantity $\Phi_0$ be the electric potential.\footnote{We will add a subscript zero with the physical quantities (\emph{e.g.} $M_0, T_0 \dots $) for the Einstein-Hilbert action \emph{i.e.} leading order contribution.} Additionally, the cosmological constant's conjugate quantity, $\Theta_0$, can be understood geometrically as the proper volume weighted by the killing vector's norm \cite{Kastor:2009wy, Cvetic:2010jb, Jacobson:2018ahi}
\begin{eqnarray}\label{eq:thetadef}
	\Theta_0 = \int_{\text{BH}} |\xi| \ dV - \int_{\text{AdS}} |\xi|\ dV \,,
\end{eqnarray}
where $|\xi|$ is the norm of the killing vector, the integration will take place over the constant time hypersurface of the black hole and pure AdS spacetime. The generalised Smarr relation for $d+1$ dimensional bulk is given by \cite{Smarr:1972kt, Banerjee:2010ye}
\begin{eqnarray}\label{eq:smarr}
	M_0 = \frac{d-1}{d-2} \frac{T_0 A_0}{4 G_N} + \Phi_0 Q_0 - \frac{1}{d-2}\frac{\Theta_0 \Lambda_0}{4 \pi G_N}.
\end{eqnarray}
In extended thermodynamics, the bulk pressure is identified as the cosmological constant
\begin{equation}\label{eq:Pdef}
	P_0 = - \frac{\Lambda_0}{8\pi G_N}, \quad \text{with} \quad \Lambda_0 = - \frac{d(d-1)}{2 L_0^2} \,,
\end{equation}
where $L_0$ is the AdS curvature radius. Without considering $G_N$ fluctuation, the first law (\ref{eq:bulkfirstlaw}) and generalised Smarr relation \eqref{eq:smarr} takes the form 
\begin{align}
	d M_0 &= T_0 dS_0 + V_0 dP + \Phi_0 dQ_0 \,, \label{eq:bulkfirstlawnoG}\\
	M_0 & = \frac{d-1}{d-2} T_0 S_0 + \Phi_0 Q_0 - \frac{2}{d-2} P_0 V_0  \,.\label{eq:Smarrrel}
\end{align}
However, there is some disagreement regarding the bulk first law, dual to the first law of thermodynamics, in boundary field theory. Firstly, the energy of the boundary theory is dual to the ADM mass $M_0$ of the black hole, whereas in extended thermodynamics $M_0$ is identified as the thermodynamics enthalpy $H$ rather than the internal energy of the black hole, \cite{Kubiznak:2012wp, Kubiznak:2016BH-chem-lambda}. Secondly, one can compute the asymptotic stress tensor and the pressure of the boundary theory \cite{Balasubramanian:1999re, Balasubramanian:1998sn}. This approach of evaluating the boundary pressure does not yield the bulk pressure specified above, and the spatial volume of the CFT $\mathcal{V} \sim L_0^{d-1}$ is not related to the thermodynamic volume $V_0$ of the black holes \emph{i.e.} the conjugate quantity of $\Lambda_0$. As a result, the boundary's first law is unable to be interpreted precisely as the bulk first law \eqref{eq:bulkfirstlawnoG}.

It is possible to address the discrepancy in the thermodynamics variable on both sides by choosing additional thermodynamic variables. With this new set of thermal variables, we can have a one-to-one map between the extended black hole thermodynamics in bulk and the thermodynamics variable of the boundary CFT. In the holographic dictionary, the Einstein gravity with AdS curvature length $L_0$ and effective Newton's constant $G_N$ in $d+1$ dimensional contains a dual central charge of the CFT theory presented as $ \mathtt{c} \sim L_0^{d-1}/G_N$; hence the variation of cosmological constant $\Lambda_0 \sim 1/L_0^2$ will lead to the variation of \emph{central charge} $\mathtt{c}$ of boundary CFT or the number of colors $N$ in the dual gauge theory as well as the variation of Newton's constant $G_N$ in the gravity theory \emph{i.e.} the bulk thermodynamics as a ``bookkeeping'' device \cite{Karch:2015rpa, Visser:2021eqk, Cong:2021jgb, Cong:2021fnf}. The holographic interpretation of charged AdS black holes is intriguing in this new paradigm where Newton's constant is the dynamics parameter. Central charge criticality studies for non-linear electromagnetic black holes \cite{Kumar:2022fyq, Bai:2022vmx}, Gauss-Bonnet black holes \cite{Kumar:2022afq, Qu:2022nrt}, and other types of black holes studies \cite{Gong:2023ywu, Zhang:2023uay} have all shown that this transition is determined by the degrees of freedom of its dual field theory in the large N limit.

As a consequence of the $G_N$ variation, the first law \eqref{eq:bulkfirstlaw} can be reformulated in the following manner
\begin{equation}\label{eq:bulkfirstlaw2}
	\begin{aligned}
		d M_0 = T_0 d S_0 + \frac{\Phi_0}{L_0} d (L_0 Q_0) - \frac{M_0}{d-1}\frac{d L_0^{d-1}}{L^{d-1}} + \Big(M_0 - T_0 S_0 - \Phi_0 Q_0\Big) \frac{d(L_0^{d-1}/G_N)}{L_0^{d-1}/G_N} \,,
	\end{aligned}
\end{equation}
so that it can be directly mapped to the first law of the boundary theory \cite{Visser:2021eqk, Cong:2021fnf, Cong:2021jgb}. Here the first two terms are analogous to the boundary first law, and $L_0^{d-1}$ is proportional to the thermodynamic volume of the boundary theory. Thus the coefficient of $d L_0^{d-1}$ term is accordingly identified with the pressure of the boundary theory, which satisfies the equation of state: $ M_0 = E = (d-1) p \mathcal{V}$. Finally, the last term in the first law $d(L_0^{d-1}/G_N)/(L_0^{d-1}/G_N)$ is identified with the variation of the central charge $c$, and its coefficient is a new chemical potential $\mu_c$.\footnote{In \cite{Ahmed:2023snm}, an alternate formulation to write the holographic first law is proposed, which is precisely dual to the first law of the extended blackhole thermodynamics by treating the conformal factor of the AdS boundary as a thermodynamic parameter and allowing the AdS radius and the CFT volume to vary independently. \label{foot3}} This new chemical potential  satisfies the boundary Euler relation
\begin{equation}\label{eq:euler}
	E = M_0 = T_0 S_0 + \Phi_0 Q_0 + \mu_c \mathtt{c}.
\end{equation}

The discussion of the effect of higher derivative terms on the boundary first law provides an intriguing glimpse into the thermodynamics of boundary field theory  \cite{Dutta:2022wbh}. Thereafter, the charged black hole in the presence of these higher derivative terms and their holographic dual phases are explored in  \cite{Kumar:2022afq, Qu:2022nrt}. Nevertheless, the anomalous contribution of higher derivative terms to the first law of boundary theory is absent. This work concentrates on answering these questions and hence divided into two parts: in the first, we have generalized the findings of  \cite{Dutta:2022wbh} to a charged black hole system, and in the second, we investigate how black hole charge affects the black hole thermodynamics in higher derivative theories and their holographic dual theory. In the boundary theory, we investigate the influence of the black hole charge on the recently revealed phase behaviour \cite{Visser:2021eqk, Cong:2021jgb, Cong:2021fnf} of the dual to the higher derivative corrected charged black hole.

Following the similar techniques as \cite{Dutta:2022wbh}, this paper establishes the connection between the bulk and the boundary thermodynamics in a generic 4-derivative theory of gravity coupled to $U(1)$ gauge field in the presence of the cosmological constant. Firstly, we present the general structure because the action of higher derivative terms modifies the geometry and charges of black holes, so the first law of thermodynamics is also modified. Extending the connection over the supergravity limit may appear inconsequential due to higher derivative modifications to all thermodynamic variables. But one has to be careful because another holographic computation another central charge $\mathtt{a}$ as the presence of higher derivative terms and both central charges can be expressed in the dimensionless quantities $L_0^3/G_N$ and $\alpha'/L_0^2$ in the gravitational side, which implies that the variation of the cosmological constant in bulk also induces a variation of multiple central charges at the boundary theory. To disentangle the variation of $\mathtt{c}$ and $\mathtt{a}$ from the variation of $N$ and volume $V$, one needs to include the variation of $\alpha'$ (along with $G_N$ and $L$) in bulk as a bookkeeping device. We show that by identifying the appropriate thermodynamic variables as $\mathtt{c}_+$ and $\mathtt{c}_-$, the bulk first law is naturally interpreted as the boundary first law, and the bulk Smarr relation generates the generic Euler relation of the boundary theory. To support our generic result, we put some examples of charged black holes in the presence of higher derivative theories, where one is a charged Gauss-Bonnet AdS black hole, and another is a charged AdS black hole in generic 4-derivative gravity.

Next, we investigate the phase behaviour of the charged black hole in the higher derivative gravity theory after establishing a one-to-one correlation between bulk and boundary thermal parameter space. The dimension of the thermodynamic phase will grow due to the different chemical potentials endowed in the phase structure of charged AdS black holes with higher derivative terms. After that, we study the critical behaviour analysis of the corrected black holes, where we compute the critical value of all the thermal quantities like temperature, pressure and central charge up to $\mathcal{O}(\alpha ')^2$. To investigate the critical phenomenon of the corrected charged black hole, we examine the plot of the free energy of black holes w.r.t. their respective Hawking temperature in canonical ensemble and examine the relevant phase behaviours by varying the various parameters. Further probing the Gauss-Bonnet black hole's extended phase space, we study the chemical potentials' $\mu_{\pm}$ critical behaviour conjugate to the new variable $\mathtt{c}_{\pm}$ in boundary field theory. We noticed that $\mu_+$ behaves in a swallowtail format corresponding to different phases in the boundary theory that is dual to small black hole, large black hole and the unstable branch in bulk. Furthermore, $\mu_+$ behaves in a manner similar to how the chemical potential $\mathcal{A}_{\alpha '}$ emerges in the presence of higher derivatives. Besides, we do the identical analysis for charged black holes in generic 4-derivative gravity.

Summarizing this investigation, we delve into the complexities of the expanded thermodynamics associated with the charged black hole, considering the effect of the higher derivative term (in particular the 4-derivative gravity). However, the distinctive contribution of our study lies in its emphasis on elucidating the holographic measurement of the extended thermodynamic properties exhibited by charged black holes within the context of higher derivative theories. Our research seeks to unveil the unique holographic facets manifest when higher derivative terms are introduced into the analysis of charged black hole thermodynamics, thereby enriching our comprehension of the interplay between gravitational physics and holography in this particular context. Additionally, we delve into an in-depth examination of the distinct phases of Conformal Field Theory (CFT) that correspond to the $\alpha'$-corrected black hole in the bulk, which entails a meticulous exploration of the intricate interplay between the properties of the black hole and the corresponding features within the dual Conformal Field Theory. By scrutinizing these various phases, we aim to unravel the nuanced relationships and implications arising from the $\alpha'$ corrections, shedding light on the profound connections between the microscopic and macroscopic descriptions of the gravitational system.

The plan of this paper is the following. In Section \ref{sec:firstlaw}, we discuss the bulk first law and the Smarr relation of the $U(1)$ charged black hole in the presence of generic higher derivative terms. After that, in Section \ref{sec:bdyfirstlaw}, we demonstrate how to write the holographic dual of bulk first law with higher derivative terms \textit{i.e.} we study the correspondence between the bulk first law and the boundary first law. In Section \ref{sec:HDexamp}, we discuss a few examples to support our generic results, which contain charged black hole Gauss-Bonnet gravity followed by charged black holes in generic 4-derivative gravity. In Section \ref{sec:phasestr}, we discuss the phase structure of these black holes. Eventually, with some final remarks and open questions, we conclude our results in Section \ref{sec:conclusion}.


\section{Extended thermodynamics of higher-derivative corrected black hole}\label{sec:firstlaw}

In this section, we compute the Smarr relation and the first law of black hole thermodynamics with higher derivative terms. In the throat limit \cite{Maldacena:1997re}, the effective $d+1$ dimensional action has the following approximate form
\begin{equation}\label{eq:actionGen}
	\begin{aligned}
		\mathcal{S} = \frac{1}{16 \pi G_N} \int d^{d+1}x \sqrt{-g} \bigg( R - 2\Lambda_0 - \frac{1}{4} F_{\mu\nu} F^{\mu\nu} + \sum_{n \geq 1}\left( \alpha' \right)^{n} \mathcal{L}_{HD} \left( R^{n+1},F^{2n + 2} \right) \bigg) \,,
	\end{aligned}
\end{equation}
where $\alpha'$ is proportional to the square of the string length, and  $\mathcal{L}_{HD}$ is the Lagrangian density entangling the broad set of higher derivative terms arising from the contraction of the curvature tensor and electromagnetic field strength. Such higher derivative terms emerge in low energy effective action of different closed string theories and the structures of these higher derivative terms are completely fixed for a specific string theory\footnote{For example, the appearance of curvature square term \emph{e.g.} $ \textbf{Riemann}^2 $ in heterotic string theory is well known and $ \textbf{Riemann}^4 $ terms appear in superstring theories whereas $ \textbf{Riemann}^3 $ appears in bosonic string theory.}. We assume that $\alpha' \mathcal{R} \ll 1$, where $\mathcal{R}$ implies the curvature scale of the solution. The only parameter which appears in front of these terms is different powers of $\alpha'$ as this is the only dimension full parameter in the theory. In the supergravity limit $\alpha' \to 0 $, all the higher derivative terms drop out. 

As $\alpha' \mathcal{R} \ll 1$, we perturbatively solve Einstein's equation and Maxwell's equation for higher derivative theories, and as a result of those higher derivative terms, all the thermal quantities related to the black hole, such as the black hole temperature, entropy, ADM mass, and the chemical potential conjugate to the electric charge of the black hole, receive correction, for detailed computation we refer to the section \ref{sec:HDgravity}. Before we constructed the modified first law of thermodynamics for higher derivative theories, we first note the action \eqref{eq:actionGen} admits a vacuum AdS solution such that $R = 20/L^2$, where $L$ be the effective AdS length \cite{Cremonini:2009ih, Cremonini:2019wdk}. For general higher derivative theory, the effective AdS length can be written as $L = L_0 \mathsf{L}(\alpha'/L_0^2) $ \footnote{Where $\mathsf{L}(\alpha'/L_0^2)$ depends on the form of the higher derivative term present in theory.} (for example, Eq. \eqref{eq:LcdefHD}) and the vacuum metric form in asymptotic AdS geometry with $L$ is
\begin{equation}\label{eq:vacmet}
	ds^2 \sim \left(1+\frac{r^2}{L^2}\right) dt^2 + \left(1+\frac{r^2}{L^2}\right)^{-1} dr^2 + r^2 d\Omega_3^2 \,.
\end{equation}
where $d\Omega_3^2$ be the metric on unit 3-sphere. As a result, denoting the effective radius by $L$, the corrected the cosmological constant $\Lambda$ can be defined as \footnote{One can proceed with the modified definition of thermodynamics pressure where $ P = - \Lambda / (8 \pi G_N) $ instead of \eqref{eq:Pdef}, then we have to keep in mind that our asymptotic structure modifies accordingly, and consequently, the thermodynamic quantities like thermodynamic volume etc. as discussed in appendix \ref{app:GBthermalqu}.} 
\begin{equation}\label{eq:lamdacorrdef}
	\Lambda_0 \longrightarrow {\Lambda} = - \frac{d(d-1)}{2 {L}^2}.
\end{equation}

The black hole thermodynamic quantities are related to each other by the Smarr relation. In the context of higher derivative theories the Smarr relation is well studied, which includes the Gauss-Bonnet black hole \cite{Wei:2012ui, Zou:2014mha}, $f(R)$ gravity theory \cite{Chen:2013ce}, for Lovelock gravity in AdS spacetime \cite{Kastor:2010gq, Sinamuli:2017rhp}, the quasitopological gravity \cite{Hennigar:2015esa}, Einstein-cubic gravity \cite{Hennigar:2016gkm} and other higher curvature gravity theories \cite{Castro:2013pqa, Sinamuli:2017rhp}.
The mass of the black holes plays a central role in understanding the thermodynamics and behavior of black holes. Apparently, in theories that involve higher derivatives, the mass of a black hole can be described in a general way as follows:
\begin{align}
	M=M(A, G_N, Q, \Lambda_0,\alpha') .
\end{align}
Taking variation of both side and incorporating the variation of gravitational constant $G_N$ and the variation of coupling constant of higher derivative term, the first law turns out to be
\begin{align}\label{eq:firstlawHD}
	d M = \frac{\kappa}{8 \pi G_N} dA + \Phi dQ + \frac{\Theta}{8 \pi G_N} d\Lambda_0 	- \Big(M - \Phi Q \Big) \frac{dG_N}{G_N} + \frac{\mathcal{A}_{\alpha '}}{G_N} d \alpha ' \,.
\end{align}
Following the scaling argument from \cite{Kastor:2009wy, Kastor:2010gq, Kubiznak:2016qmn}, the Smarr relation in extended phase space (including $P$ and $V$ variable) with generic higher derivative correction\footnote{In this article, we will add a subscript zero (\emph{e.g.} $M_0, T_0 \dots $) to represent the physical quantities that do receive any $\alpha '$ corrections while the upper case letters $ ( M, T, S, \Phi, Q,  \dots)$ without any subscript will denote that physical quantities which receive the $\alpha '$ corrections.} in $(d+1)$-dimension arranged in a straightforward form as
\begin{equation}\label{eq:smarrHD}
	{M} = \frac{d-1}{d-2} \frac{\kappa A}{8 \pi G_N} - \frac{1}{d-2} \frac{\Theta  \Lambda_0}{4 \pi G_N} + \Phi \, Q + \frac{2}{d-2} \frac{\mathcal{A}_{\alpha'}}{G_N} \alpha ' \,.
\end{equation}
The quantity $A$ appearing in the first term in Smarr relation is given by $ A = 4 G S_{\text{Wald}} $, where $S_{\text{Wald}}$ be the Wald entropy as defined in \eqref{eq:Wentdef}. We call this quantity the ``Wald area''. Later, we shall see that the horizon area remains unchanged under higher derivative corrections in our parametrization. The Wald area equals the horizon area in the $\alpha' \to 0$ limit. 
Here the new thermodynamics variable $ \mathcal{A}_{\alpha'} $ be the chemical potential conjugate to $\alpha'$ \cite{Kubiznak:2016BH-chem-lambda, Sinamuli:2017rhp, Hennigar:2015esa, Zou:2014mha, Kumar:2022afq, Qu:2022nrt, Wei:2012ui, Chen:2013ce, Dutta:2022wbh} and $ \mathcal{A}_{\alpha'} d \alpha'$ term in first law will disappear in the supergravity limit $(\alpha' \to 0)$. Another important point to note here is that, unlike two derivative gravity, the variable $\Theta$ does not have the geometrical meaning anymore; it's just conjugate quantity corresponds to cosmological constant and is defined as the generalization of \eqref{eq:thetadef}. 

We can restore the first law and Smarr relation for the RN-AdS-BH as given in \eqref{eq:bulkfirstlaw} and \eqref{eq:smarr} respectively in the limit $ \alpha' \to 0$. The first law and Smarr relation are compatible with \cite{Kastor:2010gq, Dutta:2022wbh} in the limit $Q \to 0$. Next, by choosing the appropriate boundary thermodynamic variables, we will untangle the boundary first law from the bulk first law by exploiting the AdS/CFT dictionary. In this setup, the bulk Smarr relation reduces to the generic Euler relation of boundary CFT.


\section{Holographic first law}\label{sec:bdyfirstlaw}

The holographic interpretation of the extended black hole has been unclear for many years, and multiple frames of reference suggested earlier \cite{Johnson:2014yja, Kastor:2014dra, Karch:2015rpa, Dolan:2016jjc} since it is not straightforward to map the bulk first law to the boundary first law. The mapping of the $T dS $ or $\Phi dQ $ term is evident from the beginning because one can map the Hawking temperature to the temperature of field theory, and the same goes with the conserved charge of the bulk and boundary theory. But the presence of the $ V dP $ term complicates the mapping because, on the other side, the CFT volume is proportional to $\sim L_0^{d-1}$ and the holographic CFT dual to Einstein gravity has a dictionary $\mathtt{c} \sim L_0^{d-1}/G_N$. Thus the $V dP$ term leads to a degeneracy as it induces $-p \, d\mathcal{V}$ and $\mu \, d\mathtt{c}$ terms in the boundary first law, which are not self-reliant. As discussed in Section \ref{sec:intro},  we need the variation of Newton's constant to disentangle this degeneracy. As discussed in footnote \ref{foot3}, an alternative proposal to formulate the holographic first law is proposed such that the dual to the first law of extended black hole thermodynamics is derived by treating the conformal factor of the AdS boundary as a thermodynamic parameter and allowing the AdS radius and the CFT volume to vary independently. The variation of the bulk cosmological constant corresponds to changing the CFT central charge and the CFT volume \cite{Ahmed:2023snm}.

The presence of the higher dimensional operators with coupling constant $\alpha'$ uplifts the conformal anomaly in the boundary theory such that the $\mathcal{O}(\mathtt{c}-\mathtt{a})$ is nonzero, where the $ \mathtt{c} $ and $ \mathtt{a} $ are the anomaly coefficient known as central charge of boundary conformal field theory. We can see this from the computation of the expectation value of the CFT stress tensor given as $\left< {T^\mu}_\mu \right> = -\mathtt{a} E_4 - \mathtt{c} I_4 $ where $E_4$ is the Euler topological density, and $I_4$ be the Weyl squared term. Using the holographic renormalisation procedure, we can compute the anomaly coefficient by following \cite{Henningson:1998gx, Banerjee:2009fm}, and in the limit $\alpha' \to 0 $ we have $\mathtt{c} = \mathtt{a}  \sim L_0^{d-1}/G_N$. However, these coefficients aren't equivalent in the presence of the higher derivative correction; they disagree at $\mathcal{O}(\alpha'/L^2)$ and higher order of $\alpha'/L^2$ \cite{Nojiri:1999mh, Blau:1999vz,Banerjee:2009fm}. Thus, after incorporating the higher-derivative coupling parameter $\alpha'$ in the bulk first law, we also need an additional parameter on the boundary side. We present the inclusion of the other central charge $\mathtt{a}$ to uncover the one-to-one map between the bulk and boundary thermal parameters. However, instead of writing the first law in terms of $(\mathtt{c}, \mathtt{a})$ we define a new set
\begin{equation}\label{eq:cpmdef}
	\mathtt{c_{\pm}} = \frac{\mathtt{c} \pm \mathtt{a}}{2} \,.
\end{equation}
and we compose the boundary first law in terms of $\mathtt{c_{\pm}}$ basis such that in the limit $\alpha' \to 0 $ we get back the two derivative results. 

It is well established that for $SU(N)$ gauge theories with conformal symmetry, the central charge scales as $ \mathtt{c} \sim N^2$ at large $N$, therefore high-energy states satisfy $E \sim \mathtt{c}$ \cite{Gross:1980he, Visser:2021eqk, Witten:1998qj} and in finite though large $N$ theories the multiple central charges appear as same scale $\mathtt{c} \sim \mathtt{a}$ \cite{Banerjee:2009fm, Nojiri:1999mh, Blau:1999vz}. Hence, by definition, the internal energy of an equilibrium state depends on extensive quantities, such as entropy $S$, volume $V$, conserved charge $\tilde{Q}$, and as mentioned above for large $N$ gauge theories, it furthermore depends on the (intensive) central charges $\mathtt{c} \, \& \, \mathtt{a}$ \textit{i.e.} $E = E(S, V, Q, \mathtt{c}_+, \mathtt{c}_-)$ \cite{Cong:2021fnf,Visser:2021eqk}. We can vary the energy with respect to each quantity, which gives the first law of thermodynamics
\begin{align}\label{eq:bdyflaw}
	dE = TdS - pd\mathcal{V} + \tilde{\Phi} d\tilde{Q} + \mu_+ d\mathtt{c_+} + \mu_- d\mathtt{c_-} \,,
\end{align}
where the temperature $T$, pressure $p$, chemical potential $\tilde{\Phi}$ conjugate to $\tilde{Q}$ and the chemical potential $\mu_{\pm}$ conjugate to $\mathtt{c_{\pm}}$ are defined as
\begin{align}
	\begin{aligned}
		T \equiv \left( \frac{\partial E}{\partial S}\right)_{\mathcal{V},\tilde{Q},\mathtt{c_{\pm}}} \,, \qquad p \equiv - \left( \frac{\partial E}{\partial \mathcal{V}}\right)_{S,\tilde{Q},\mathtt{c_{\pm}}} \,, \\
		\tilde{\Phi} \equiv \left( \frac{\partial E}{\partial \tilde{Q}}\right)_{S,\mathcal{V},\mathtt{c_{\pm}}} \,, \qquad \mu_{\pm} \equiv \left( \frac{\partial E}{\partial \mathtt{c_{\pm}}}\right)_{S, \mathcal{V},\tilde{Q}} \,.
	\end{aligned}
\end{align}
The thermal quantities as scaled as 
\begin{align}
	\begin{aligned}
		& [E] = [T]  = [\mu_+] = [\mu_-]= L^{-1} \,; \qquad [\mathcal{V}] = L^{d-1} \\
		& [S] = [\tilde{Q}] = [\mathtt{c}_+] = [\mathtt{c}_-] = L^0 \,
	\end{aligned}
\end{align}
For high-energy states, the central charge, $ S, Q \sim \mathtt{c}_{\pm}$, scales with the entropy and conserved quantities, corresponding to the contribution from all degrees of freedom. Thus, scaling of the energy function follows the scaling relation as $E( \alpha S, \mathcal{V}, \alpha \tilde{Q}, \alpha \mathtt{c}_+, \alpha \mathtt{c}_-) = \alpha E( S, \mathcal{V}, \tilde{Q}, \mathtt{c}_+, \mathtt{c}_-)$, where $\alpha$ be some dimensionless scaling parameter. Differentiating with respect to $\alpha$ and putting $\alpha = 1$ leads to the Euler equation
\begin{align}\label{eq:eulerHD}
	E = T S + \tilde{\Phi} \tilde{Q} + \mu_+ \mathtt{c}_+ +  \mu_- \mathtt{c}_- \,.
\end{align}
Since the volume does not generally scale as $\mathtt{c}_{\pm}$, we discover that pressure and volume are absent from this Euler equation. As shown in \cite{Visser:2021eqk, Cong:2021fnf}, another scaling argument gives rise to the equation of state for gauge theories
\begin{align}
	E = (d-1)p \mathcal{V}
\end{align}
Now we will show how the bulk first law \eqref{eq:firstlawHD} and the boundary first law \eqref{eq:bdyflaw} are related in generic higher derivative theories and how Smarr relation \eqref{eq:smarrHD} for black hole reduces to Euler relation \eqref{eq:eulerHD} for boundary gauge theory.

From the dimensional analysis, the generic form of $\mathtt{c_{\pm}}$ can be written as 
\begin{align}\label{eq:cpmgeneric}
	\begin{aligned}
		\mathtt{c}_+ & = \frac{L_0^{d-1}}{128 \pi G_N} \, {h}_+(\alpha'/L_0^2)\,, \quad \text{and}
		\\ \mathtt{c}_- & = \frac{L_0^{d-1}}{128 \pi G_N} \, h_-(\alpha'/L_0^2) \,,
	\end{aligned}
\end{align}
where $h_+$ and $h_-$ are functions of dimensionless parameter $\alpha'/L_0^2$ and depend on the nature of the higher derivative terms added in theory. In holographic theory, they also satisfy
\begin{eqnarray}
	\begin{aligned}
		& h_+(\alpha'/L_0^2) \sim \mathcal{O}(1) \quad \text{and} \\
		& h_{-}(\alpha'/L_0^2) \sim \mathcal{O}\left(\frac{\alpha '}{L_0^2}\right) \quad \alpha' \to 0.
	\end{aligned}
\end{eqnarray}
Varying equations (\ref{eq:cpmgeneric}) we find
\begin{equation}\label{eq:dcpm}
	\frac{d\mathtt{c}_\pm}{\mathtt{c}_\pm } = \frac{dL_0^{d-1}}{L_0^{d-1}} - \frac{dG_N}{G_N} + \frac{h'_\pm}{h_\pm } d \left(\frac{\alpha'}{L_0^2}\right) \,, 
\end{equation}
from \eqref{eq:dcpm}, we replace $d\alpha'$  and $dG_N$ in \eqref{eq:firstlawHD} in terms of $d \mathtt{c_{\pm}}$ in the bulk first law. We also use the Smarr relation\eqref{eq:Smarrrel} to replace $\Theta$ in the first law. After simplification the final result is given by\footnote{The standard holographic dictionary as $E = M, \tilde{\Phi} = \Phi/L_0, \tilde{Q} = Q L_0 $ for the CFT dwelling on the boundary of AdS spacetime with the AdS length $L_0$.}
\begin{align}\label{eq:bdyfirstlaw}
	dM = T dS + {\tilde{\Phi}} d {\tilde{Q}}- \left(\frac{M}{d-1}\right) \frac{dL_0^{d-1}}{L_0^{d-1}}  + \bigg( \frac{ h_-' \left(M - T S - \tilde{\Phi} \tilde{Q} \right) - h_- \mathcal{A}_{\alpha'} L_0^2}{ \left(\mathtt{c}_+ h_-' - \mathtt{c}_- h_+' \right)}\bigg) d\mathtt{c}_+ \nonumber \\
	+ \bigg( \frac{h_+ \mathcal{A}_{\alpha'} L_0^2 -  h_+' \left(M - S T - \tilde{\Phi} \tilde{Q} \right)}{\left( \mathtt{c}_+ h_-' - \mathtt{c}_- h_+' \right)}\bigg) d\mathtt{c}_-.
\end{align}
The coefficient of $d\mathtt{c}_\pm$ denoted as the conjugate chemical potentials $\mu_\pm$ such that
\begin{align}\label{eq:mupmexp}
	\begin{aligned}
		\mu_+ & = \left( \frac{ h_-' \left(M - T S - \tilde{\Phi} \tilde{Q} \right) - h_-\mathcal{A}_{\alpha'} L_0^2 }{ \left(\mathtt{c}_+ h_-' - \mathtt{c}_- h_+' \right)}\right) \,, \\
		\mu_- & =  \left( \frac{h_+ \mathcal{A}_{\alpha'} L_0^2 -  h_+' \left(M - S T - \tilde{\Phi} \tilde{Q} \right)}{\left( \mathtt{c}_+ h_-' - \mathtt{c}_- h_+' \right)}\right) \,,
	\end{aligned} 
\end{align}
Solving \eqref{eq:mupmexp} for $M=E$ satisfy the generic Euler relation. Alternatively if we start with boundary first law \eqref{eq:bdyflaw} and insert the value of $d \mathtt{c_{\pm}}$ from \eqref{eq:dcpm} and use generic Euler relation to simplify the relation along with $-2 dL_0/L_0 = d\Lambda_0 / \Lambda_0$, we get the first law as
\begin{align}
	\begin{aligned}
		dM = &  T d\left(\frac{A}{4G_N}\right) + \Phi dQ - \left(M - \Phi Q\right) \frac{dG_N}{G_N}  + \frac{L_0^{d-1}}{G_N} \left( \mu_+ h_+' + \mu_- h_-'\right) d\alpha'  \\
		& + \bigg(\frac{8 \pi G L_0^2((d-2)(M - \Phi Q) - (d-1)TS)}{d(d-1)} - \frac{16 \pi L_0^{d-1} \alpha ' (\mu_+ h_+' + \mu_- h_-')}{d(d-1)} \bigg) \frac{d \Lambda_0}{8 \pi G_N} \,.
	\end{aligned}
\end{align}
Thus the coefficient of $d\alpha'$ is given by
\begin{align}
	\mathcal{A}_{\alpha'} =  \frac{L_0^{d-3}}{128 \pi} \left( \mu_+ h_+' + \mu_- h_-'\right) 
\end{align}
and identify the coefficient of $d\Lambda_0/8\pi G_N$ as $\Theta = - V$ gives the generalised Smarr relation given in \eqref{eq:smarrHD}. Thus we see that the bulk first law \eqref{eq:firstlawHD} can directly be identified with the extended first law of the boundary field theory \eqref{eq:bdyfirstlaw}, and the generic Smarr relation \eqref{eq:smarrHD} generates the Euler relation \eqref{eq:eulerHD}. As a consistency check, we notice that $ \mathtt{c}_+ = \mathtt{c} $ and $ \mathtt{c}_- = 0$ in the limit $ \alpha' \to 0 $ and we get back \eqref{eq:bulkfirstlaw} and \eqref{eq:smarr}.


\section{Higher derivative thermodynamics: examples}\label{sec:HDexamp}

Under a consistent truncation of string theory, higher derivative terms take a specific form and appear in the effective action of gravity theories. These terms significantly impact black hole solutions and their thermodynamics. This section shows two illustrations of charged black holes in higher derivative gravity in 5-dimension. The first example discusses the charged black hole thermodynamics associated with the Gauss-Bonnet term. The Gauss-Bonnet term was first presented by Lovelock in \cite{Lovelock:1971yv} as a natural generalization of Einstein's theory of general relativity. In the second example, we further generalized the four-derivative interactions interfering curvature square term and electromagnetic field strength, \textit{e.g.} terms such as ${F^\mu}_\nu {F^\nu}_\alpha {F^\alpha}_\beta {F^\beta}_\mu$ and $R_{\mu\nu\rho\sigma} F^{\mu\nu} F^{\rho\sigma}$, the complete higher derivative Lagrangian density illustrated in \eqref{eq:LagHDT}.

\vspace{1em}
\subsection{Example 1: Charged black holes in Gauss-Bonnet gravity}\label{sec:GBgravity}
This section discusses the electrically charged black hole in Gauss-Bonnet gravity with the cosmological constant \cite{Cai:1998vy, Cai:2001dz, Cai:2013qga, Cvetic:2001bk, Nojiri:2001aj}. In contrast, the Gauss-Bonnet gravity is an extension of Einstein's gravity with the Euler topological term in the domain of Lovelock gravity, whereas the Lovelock theory contains the sum of extended Euler densities. In contrast, the Lovelock gravity theory has some incredible features among the gravity theory with higher derivative curvature terms \cite{Boulware:1985wk}. Alternatively, Gauss-Bonnet gravity becomes interesting because string theory anticipated such theories. The Calabi-Yau three-fold compactification of M-theory presents the effective Einstein-Gauss-Bonnet theory with a suitable choice of coefficient of the Gauss-Bonnet term corresponding to Lovelock theory \cite{Ferrara:1996hh, Fernandes:2022zrq}. The one-loop effective action of heterotic string theory in the Einstein frame exhibits Gauss-Bonnet terms with the coupling constant of form $\alpha ' e^{\varphi}$ \cite{Nepomechie1985, Zwiebach1985315, Callan198678, Fernandes:2022zrq} (where $\varphi$ is a dynamical scalar field known as the dilaton).
The Maxwell-Gauss-Bonnet action (in the Einstein frame) with the negative cosmological constant $(\Lambda)$ in  $5$-dimension is given by
\begin{equation}\label{eq:actionGB}
	\begin{aligned}
		S = \frac{1}{16 \pi G_N} \int_{\mathcal{M}} d^{5}x \sqrt{-g} \bigg[ R - 2 \Lambda_0 - \frac{1}{4} F_{\mu\nu} F^{\mu\nu} + \alpha ' \Big( R^2 - 4 R_{\mu\nu} R^{\mu\nu} 
		+ R_{\mu\nu\rho\sigma} R^{\mu\nu\rho\sigma}  \Big) \bigg] \,,
	\end{aligned}
\end{equation}
where $\alpha '$ is proportional to the square of string length and is positive in heterotic string theory and $ \Lambda_0 = - 6 / L_0^2 $ where $L_0$ be the AdS curvature length \footnote{Here $L$ be the effective AdS length $L = L_{\text{eff}}$. One can compute the $L_{\text{eff}}$  from the asymptotic structure of the metric or the Ricci scalar, the effective AdS length ig given by $$ L^2 = \frac{L_0}{2} \left( L_0 + \sqrt{ L_0^2 - 8 \alpha '} \right)\,,$$  with the effective AdS length $L$ and the AdS length $L_0$ \emph{i.e.} the uncorrected AdS length, we can present the boundary thermal quantities in therm of either $L$ or $L_0$, given the convenience of simplified expression. \label{foot:LGBdef}}. The Gibbons-Hawking-York (GHY) term is required for a well-defined variation principle with respect to the metric; for the Gauss-Bonnet correction, the GHY term has the following form:
\begin{align}\label{eq:bdyactGB}
	\begin{aligned}
		S^{(1)}_{\text{bdy}} = \frac{1}{8 \pi G_N}\int d^4 y \sqrt{-\gamma} \Bigg[ K + \alpha ' \Bigg(-\frac{2}{3}K^3+2KK_{ab}K^{ab}	-\frac{4}{3}K_{ab}K^{bc}K_c^a \\ -4\left(\mathcal{R}_{ab}-\frac{1}{2}\mathcal{R}h_{ab}\right)K^{ab}\Bigg) \Bigg] \,,
	\end{aligned}
\end{align}
where $\mathcal{R}_{ab}$ is the Ricci tensor at the boundary, $K$ be the trace of the second fundamental form \emph{i.e.} the extrinsic curvature tensor, which is the measure of how the normal to the hypersurface changes and $\gamma_{ab}$ is the induced metric on the hypersurface. A well-defined variational w.r.t. gauge field required a boundary term given by
\begin{align}\label{eq:bdyactMax}
	\begin{aligned}
		S^{(2)}_{\text{bdy}} = \frac{1}{16\pi G_N} \int_{\partial \mathcal{M}} d^{4} y \sqrt{-\gamma} & n_{\mu} F^{\mu\nu}A_{\nu} \,,
	\end{aligned}
\end{align}
which is also known as the Hawking-Ross boundary term.

\subsection*{Black Hole solution and thermodynamics quantities}
An exact solution for the Gauss-Bonnet corrected Einstein's equation can be obtained for spherically symmetric and stationary spacetime. An ansatz for the static and spherical symmetric metric is
\begin{equation}
	ds^2 = - f(r) dt^2 + \frac{1}{f(r)} dr^2 + r^2 d\Omega^2_{3} \,,
\end{equation}
where \(\Omega_{3}\) is line element of unit sphere in \(3\)-dimensions and $f(r)$ is 
\begin{equation}
	f(r) = 1 + \frac{r^2}{4 \alpha '} \left( 1 - \sqrt{1 - \frac{8 \alpha'}{L_0^2} + \frac{8 \alpha ' m}{r^4} - \frac{8 \alpha ' q^2}{r^{6}}} \right) \,.
\end{equation}
Here, $m$ is related to the ADM mass $M$ of the black hole, and the parameter $q$ is related to the total electric charge $Q$ of the black hole as
\begin{equation}\label{eq:MandQdefGB}
	M= \frac{3 \omega_{3}}{16 \pi G_N} m \, , \qquad Q=\sqrt{\frac{3}{G_N}} \frac{\omega_{3} q}{4 \pi} \quad ; \quad \omega_{3} = 2 \pi^{2} \,.
\end{equation}
As a consequence, one can compute the ADM mass of the black hole at its outer horizon, $r_+$, which has the following form:
\begin{equation}\label{eq:massGB}
	M = \frac{3 \pi}{8 G_N}\left(\frac{L_0^2 + r_+^2}{L_0^2 r_+^{2}} + \alpha + \frac{q^2}{r_+^{2}} \right) \,,
\end{equation}
and the Hawking temperature is given by
\begin{equation}\label{eq:tempGB}
	T = \frac{f'(r_+)}{4 \pi} = \frac{L_0^2 r_+^{-2} \left( r_+^{4} - q^2 \right) + 2 \; r_+^6}{2 \pi L_0^2 r_+ (r_+^2 + 4 \alpha')} \,.
\end{equation}
In the presence of these higher derivative terms, the next endeavour is to compute the correction to the Bakenstein-Hawking entropy, which can be addressed by implementing Wald's entropy computational technique \cite{Wald:1993nt, Jacobson:1993vj}
\begin{equation}\label{eq:Wentdef}
	\mathcal{S} = 2\pi\int_{S^{3}} d^{3}\Omega\; \sqrt{- \gamma} \; \frac{\partial {\cal L}}{\partial R^{\alpha\beta\gamma\delta}} \epsilon^{\alpha\beta}\epsilon^{\gamma\delta} \,,
\end{equation}
where $\epsilon^{\mu\nu}$ is the binormal and $\gamma$ is the induced metric on the horizon. The  $\frac{\partial {\cal L}}{\partial R^{\alpha\beta\gamma\delta}}$ $ \epsilon^{\alpha\beta}\epsilon^{\gamma\delta} $ is known as the Wald entropy density of the black hole. For the Gauss-Bonnet AdS black hole \eqref{eq:actionGB}, the Wald entropy tensor will take the form,
\begin{align}\label{eq:WentdenGB}
	\begin{aligned}
		\frac{\partial {\cal L}}{\partial R^{\alpha\beta\gamma\delta}} = & \frac{1}{16\pi G_N}\bigg[ \frac{1}{2} \Big( g_{\alpha  \gamma} g_{\beta  \delta} - g_{\alpha  \delta} g_{\beta  \gamma}\Big) + \alpha'  \Big( 2 R_{\alpha  \beta  \gamma  \delta} \\
		& + 2 g_{\beta  \gamma} R_{\alpha  \delta} + 2 g_{\alpha  \delta} R_{\beta  \gamma} -2 g_{\alpha  \gamma} R_{\beta  \delta} - 2 g_{\beta  \delta}  R_{\alpha  \gamma}  \\
		& + R ( g_{\alpha  \gamma  } g_{\beta  \delta} - g_{\alpha  \delta} g_{\beta  \gamma} ) \Big)\bigg] \,,
	\end{aligned}
\end{align}
and using an appropriate definition of binormal tensor by following \cite{Dutta:2006vs}, the Wald entropy of the Gauss-Bonnet AdS black hole is
\begin{equation}\label{eq:entGB}
	S_W = \frac{\omega_{3} r_+^{3}}{4 G_N} \left( 1 + \frac{6 \alpha '}{r_+^2} \right) \,.
\end{equation}

\subsection*{Holographic first law and the chemical potential}
In this section, we work with 5-dimensional spacetime. The particular justification for that is we employ holographic renormalisation to compute the anomaly coefficient, which plays a crucial role in writing the holographic first law, and the computation of these anomaly coefficients for generic spacetime is highly challenging, which is beyond the scope of this work, so we use a particular example of it \cite{Henningson:1998gx, Nojiri:1999mh, Banerjee:2010ye}. 

As discussed in Section \ref{sec:firstlaw}, the first law of thermodynamics can be computed from the variation of ADM mass of the black hole. Thus the first for charged black holes in Gauss-Bonnet theory takes the form
\begin{align}
	\begin{aligned}
		d \, {M} = \frac{T }{4 G_N} dA + \Phi dQ + \frac{\Theta}{8 \pi G_N} d\Lambda_0 - \left(M - \Phi Q \right) \frac{dG_N}{G_N}	+ \frac{\mathcal{A}_{\alpha '}}{G_N} d \alpha ' \,,
	\end{aligned}
\end{align}
where we allow the variation of Newton's constant $G_N$ along with the cosmological constant $\Lambda$ and Gauss-Bonnet parameter $\alpha'$. The geometric volume $\Theta$ and the conjugate chemical potential corresponding to the Gauss-Bonnet parameter is $\mathcal{A}_{\alpha '}$ is given by
\begin{align}
	\Theta & = -\frac{1}{2} \pi^2 r_+^4 \,, \qquad \text{and}\\
	\mathcal{A}_{\alpha '} & = \frac{3 \pi  \left(L_0^2 \left(4 q^2+4 r_+^2 \alpha '-3 r_+^4\right)-8 r_+^6\right)}{4 L_0^2 r_+^2 \left(r_+^2 + 4 \alpha '\right)} \,.
\end{align}
To uncover the bulk first law in the terms of boundary variables, we compute the anomaly
coefficients $\mathtt{c}$ and $\mathtt{a}$ in presence of the Gauss-Bonnet term in bulk theory by following \cite{Henningson:1998gx, Nojiri:1999mh, Banerjee:2010ye}. The anomaly coefficient is given by
\begin{align}\label{eq:candadefGB}
	\mathtt{c} & = \frac{L_0^3}{128 \pi G_N} \left( 1 - \frac{7 \alpha'}{ L_0^2}\right) \,, \qquad \text{and} \qquad
	\mathtt{a} & = \frac{L_0^3}{128 \pi G_N} \left( 1 - \frac{15 \alpha'}{L_0^2}\right) \,.
\end{align}
As discussed in Section \ref{sec:bdyfirstlaw}, we can replace the variation of $dG_N $ and $d\alpha '$ in terms of boundary variable as $d\mathtt{c}_{\pm}$, as a consequence of that first law of Gauss-Bonnet AdS black hole takes the form as
\begin{align} \label{eq:bdyflGB}
	d E = T dS + {\tilde{\Phi}} d{\tilde{Q}} - p d\mathcal{V} + \mu_+ d\mathtt{c}_+ + \mu_- d\mathtt{c}_- \,,
\end{align}
where $p = M / 3 \mathcal{V}$ represent the field theory pressure with boundary volume $\mathcal{V} \propto L_0^3 $ and $\mu_{\pm} $ are the chemical potentials conjugate to the new boundary variables depending on the central charge of the boundary theory, where $\mu_{\pm} $ can be represented in term of black hole parameter \footnote{Here we write the expression of $\mu_{\pm}$ in term of $L_0$, one can write it in terms of $L$ using the expression as shown in footnote \ref{foot:LGBdef}.}
\begin{align}\label{eq:mupexpGB}
	\begin{aligned}
		\mu_+ & = - \frac{16 \pi ^2 \left(L_0^2 \left(q^2 - r_+^4\right)\left(r_+^2 + 12 \alpha '\right) + 12 r_+^6 \alpha '+ r_+^8\right)}{L_0^5 \left(r_+^2 + 4
			\alpha '\right)} \,,
	\end{aligned}
\end{align}
and
\begin{align}\label{eq:mudexpGB}
	\begin{aligned}
		\mu_- & = \frac{4 \pi ^2 }{L_0^5 r_+^2 \left(r_+^2 + 4 \alpha '\right)} \Big( 12 q^2 \left(2 L^2-11 \alpha '\right) \\
		& +r_+^2 \left(24 L^2
		\alpha '-11 q^2\right)+r_+^4 \left(132 \alpha '-18
		L^2\right)-37 r_+^6\Big) \\
		& -\frac{44 \pi ^2 r_+^4 \left(r_+^2-12 \alpha '\right)}{L_0^5 \left(r_+^2 + 4 \alpha '\right)} \,,
	\end{aligned}
\end{align}
and they satisfy the Euler relation of the boundary theory as presented in \eqref{eq:eulerHD}.


\subsection{Example 2: Charged black holes in generic 4-derivative gravity}\label{sec:HDgravity}
We start with a complete set of four-derivative corrections to the Einstein-Maxwell action. In general, the 4-derivative gravity action is
\begin{eqnarray}\label{eq:HDaction}
	S_{bulk} = -\frac{1}{16 \pi G_N} \int d^{5} x\sqrt{-g} \left( \mathcal{L}_{2 \partial} + \alpha ' \;  \mathcal{L}_{4 \partial} \right) \,,
\end{eqnarray}
where $\alpha '$ is proportional to square of the string length and $\mathcal{L}_{2 \partial}$ be the standard Einstein-Maxwell Lagrangian with cosmological constant
\begin{equation}\label{eq:LagEHM}
	\mathcal{L}_{2 \partial} = R -2 \Lambda_0 -\frac{1}{4} F_{\mu\nu} F^{\mu\nu} \,,
\end{equation}
and $\mathcal{L}_{4 \partial}$ be the maximal potential contraction between gauge field and curvature tensor
\begin{align}\label{eq:LagHDT}
	\begin{aligned}
		\mathcal{L}_{4 \partial} = & b_1 R^2 + b_2 R_{\mu \nu} R^{\mu \nu} + b_3 R_{\mu \nu \rho \sigma} R^{\mu \nu \rho \sigma} + b_4 R_{\mu \nu \rho \sigma} F^{\mu \nu} F^{\rho \sigma} \\
		& +b_5\left(F_{\mu\nu} F^{\mu\nu}\right)^2 + b_6 F^4 + b_7 R F^2 + b_8 R^{\mu \nu} F_{\mu \rho} F_\nu{ }^\rho \\
		& +b_{9} \nabla_\mu F_{\nu \rho} \nabla^\mu F^{\nu \rho} + b_{10} \nabla_\rho F^{\rho \mu} \nabla_\sigma F^\sigma{ }_\mu \\
		& + b_{11} F^{\nu \rho}\left[\nabla_\mu, \nabla_\nu\right] F_\rho^\mu \,.
	\end{aligned}
\end{align}
However, many of these terms as ambiguous up to a field re-definition \cite{Myers:2009ij, Cremonini:2019wdk, Cassani:2022lrk, Bobev:2021qxx, Bobev:2022bjm} and terms involving $ \nabla_\rho F_{\mu\nu} $ can be eliminated using the leading order Maxwell's equation and the Bianchi identities \cite{Cheung:2018cwt}. We can extract the ambiguous terms from the Lagrangian density with the proper choice of field re-definition. Thus the Lagrangian density with higher derivative terms we worked \footnote{Here we didn't consider the CS-terms or the CP odd term appear in specific dimensions because those terms are not pertinent for the static, stationary and spherically symmetric black holes} with are 
\begin{align}\label{eq:LagHD}
	\mathcal{L}_{4 \partial}= \alpha_1 R_{\mu\nu\rho\sigma}R^{\mu\nu\rho\sigma}+ \alpha_2 R_{\mu\nu\rho\sigma}F^{\mu\nu}F^{\rho\sigma} + \alpha_3 (F^2)^2+ \alpha_4 F^4 \,,
\end{align}
where $F^2 = F_{\mu\nu}F^{\mu\nu}$ and $F^4 = {F^\mu}_\nu {F^\nu}_\alpha {F^\alpha}_\beta {F^\beta}_\mu$ \,.
\subsubsection*{The Gibbons-Hawking term and boundary counterterms}
We need to add the boundary terms to the action to have a well-defined variational principle on a manifold with a boundary. Variation of Einstein-Hilbert action with respect to the metric requires the boundary term if we have the Neuman boundary condition on the metric. The Gibbons-Hawking-York boundary term for Einstein-Hilbert action is 
\begin{equation}\label{eq:actGHY}
	S_{GHY} = \frac{1}{8 \pi G_N} \int_{\partial \mathcal{M}} d^4 y \sqrt{-\gamma} K \,,
\end{equation}
where $K$ be the trace of the extrinsic curvature defined as $K_{\mu\nu} = \nabla_{(\mu}n_{\nu)}$, where $n_{\mu}$ be the normal vector to the hypersurface with $\gamma_{ab}$ as the induced metric.

Now, the bulk action we worked with  is given by\footnote{This action has been explored earlier in a different context in \cite{Myers:2009ij, Cano:2022ord, Cremonini:2019wdk, Mandal:2022ztj}.}
\begin{equation}\label{eq:actionHD}
	\begin{aligned}
		S_{\text{bulk}}= & - \frac{1}{16 \pi G_N} \int_{\mathcal{M}} d^{5} x \sqrt{-g} \bigg[R - 2 \Lambda_0 -\frac{1}{4} F^2 \\
		&  + \alpha' \Big(\alpha_1 R_{\mu \nu \rho \sigma} R^{\mu \nu \rho \sigma} + \alpha_2 R_{\mu \nu \rho \sigma} F^{\mu \nu} F^{\rho \sigma} + \alpha_3\left(F^2\right)^2 + \alpha_4 F^4 \Big)\bigg] \,.
	\end{aligned}
\end{equation}
For a well-defined variation of the bulk action \eqref{eq:actionHD} with respect to the metric on spacelike or timelike boundary surfaces demands a boundary action by following the procedure in \cite{Cremonini:2009ih, Cremonini:2019wdk} is given as
\begin{align}\label{eq:bdyactHD1}
	\begin{aligned}
		S^{(1)}_{\text{bdy}} = \frac{1}{8 \pi G_N} & \int d^4 x\sqrt{-\gamma} \Bigg[ K \Big( 1 + \alpha_1 \frac{8}{L_0^2} - \alpha_1 \frac{5}{6} F^2 \Big) + 2 \alpha_1 (K n_\mu F^{\mu\lambda}n_\nu F^\nu_\lambda+K_{ab}F^{a\lambda}F^b_\lambda) \\
		& +2\alpha_2 n_\mu F^{\mu a}n_\nu F^{\nu b} K_{ab} \\
		& + \alpha_1 \Bigg(-\frac{2}{3}K^3+2KK_{ab}K^{ab}-\frac{4}{3}K_{ab}K^{bc}K_c^a -4\left(\mathcal{R}_{ab}-\frac{1}{2}\mathcal{R}h_{ab}\right)K^{ab}\Bigg) \Bigg] \,.
	\end{aligned}
\end{align}
Similarly, to have a well-behaved variation with respect to the gauge field also require the additional boundary term, which corresponds to assuming $\delta(n_\mu F^{\mu \nu}) = 0$ boundary condition from the gauge field instead of $\delta A_\mu = 0$. We found the subsequent boundary terms added to the gravitational action as the generalization of the Hawking-Ross boundary term \cite{Hawking:1995ap, Chamblin:1999tk_charge_BH} to cancel the boundary terms originating from the variation of gauge kinetic term and higher derivative operator in bulk action.
\begin{align}\label{eq:bdyactHD2}
	\begin{aligned}
		S^{(2)}_{\text{bdy}} = \frac{1}{16\pi G_N} \int_{\partial \mathcal{M}} d^{4} y \sqrt{-\gamma} n_{\mu} \bigg( F^{\mu\nu}A_{\nu}  - 4 \alpha_2 R^{\mu\nu\alpha\beta} F_{\alpha\beta} A_{\nu} - 8 \alpha _3 F^2 F^{\mu\nu}A_{\nu}\\
		 -8 \alpha_4 F^{\mu\gamma} F^{\mu\delta} F_{\gamma\delta}A_{\nu} \bigg) \,,
	\end{aligned}
\end{align}
Thus the relevant boundary terms for a well-defined variational principal for the action \eqref{eq:actionHD} is  $S_{\text{bdy}} = S^{(1)}_{\text{bdy}} + S^{(2)}_{\text{bdy}}$. If one works in the grand canonical ensemble, then the Hawking-Ross term vanishes to compute the thermodynamics quantities.

We use the holographic renormalisation procedure to extract the non-divergent part from the gravitational action computed on the background solution. This procedure entangles an appropriate boundary counterterm $S_{ct}$ to remove the divergence. Therefore the total action is presented as
\begin{align}\label{eq:totaction}
	\Gamma = S_{\text{bulk}} + S_{\text{bdy}} - S_{\text{ct}} \,.
\end{align}
To examine the appropriate counterterm \cite{Balasubramanian:1998sn, Balasubramanian:1999re}needed to regulate the action \eqref{eq:actionHD} is,
\begin{align}\label{eq:bdyctaction}
	S_{\mathrm{ct}}=\frac{1}{16 \pi G_N} \int_{\partial \mathcal{M}} d^{4} y \sqrt{-\gamma}\bigg[A+B \mathcal{R}+C_{1} \mathcal{R}^{2}
	+C_{2} \mathcal{R}_{a b}^{2}+C_{3} \mathcal{R}_{a b c d}^{2}+\cdots \bigg] \,,
\end{align}
and our foremost requirement is the identification of a vacuum AdS solution with higher-derivative term modification with $ R = - 20/ L^2 $. Where
\begin{equation}\label{eq:LcdefHD}
	L = L_0 \left(1 + \frac{\alpha_1}{3} \frac{\alpha'}{L_0^2}   \right) \,,
\end{equation}
is the corrected AdS curvature length, and the vacuum metric from asymptotic AdS geometry with $L$ is
\begin{equation}\label{eq:vacmet2}
	ds^2 \sim \left(1+\frac{r^2}{L^2}\right) dt^2 + \left(1+\frac{r^2}{L^2}\right)^{-1} dr^2 + r^2 d\Omega_3^2 \,.
\end{equation}
where $d\Omega_3^2$ be the metric on unit 3-sphere.

\subsection*{\texorpdfstring{$R^2$}{R2} corrected black hole solution in 5-dim}
Here we demonstrate the $R^2$ corrected spherical symmetric black hole solution, precisely represented by the action \eqref{eq:actionHD}. The Einstein and Maxwell equations are highly non-trivial and unsolvable with the higher derivative correction. So, we treat the higher derivative term as a perturbative correction to the Einstein-Hilbert Maxwell action and by following metric and gauge field ansatz (for a static spherically symmetric solution)
\begin{align}\label{eq:metricansatz}
	\begin{aligned}
		ds^2 & = - f(r) dt^2 + \frac{1}{g(r)} dr^2 + r^2 d\Omega^2_3 \, ,\\
		A & = A_\mu dx^\mu = \Phi(r) d t \,,
	\end{aligned}
\end{align}
where $  d\Omega^2_3 $ is a metric on a 3-sphere of unit radius. We solve the equations of motion perturbatively to obtain $f(r), g(r)$ and $\Phi(r)$. In the absence of these higher derivative terms, the equations of motion admit the Reissner–Nordstr\"{o}m black hole solution in asymptotically AdS background. The leading order metric and gauge field solution is given by
\begin{align}
	f_0(r) = g_0(r) = 1 + \frac{r^2}{L_0^2} -\frac{m}{r^2}+\frac{q^2}{4 r^4} \,,
	\Phi_0(r) = - \frac{\sqrt{3} q}{2 r} \,.
\end{align}

Now we solve the higher derivative corrected equation of motion perturbatively, and the corrected solution up to $ \mathcal{O}(\alpha ') $ takes the form
\begin{align}\label{eq:solansatz}
	f(r) = f_0(r) + \alpha ' f_1(r) \,, \nonumber \\
	\quad g(r) = g_0(r) + \alpha ' g_1(r) \,, \nonumber \\
	\Phi(r) = \Phi_0(r) + \alpha ' \Phi_1(r) \,,
\end{align}
where $f_1(r), g_1(r)$ and $\Phi_1$ stands for correction in metric and gauge field .
\begin{align}\label{eq:f1sol}
	\begin{aligned}
		f_{1} (r) = & \frac{2 \alpha _1 r^2}{3 L_0^4}-\frac{\left(13 \alpha _1+10 \alpha _2\right)q^2}{L_0^2 r^4}+\frac{2 \alpha _1 m^2}{r^6}  -\frac{2 \left(11 \alpha _1+6 \alpha _2\right) q^2}{3 r^6} \\
		& + \frac{\left(5 \alpha _1+6 \alpha _2\right) m q^2}{3 r^8} +\frac{\left(17 \alpha _1-24 \left(\alpha _2+6 \alpha _3+3 \alpha_4\right)\right) q^4}{96 r^{10}} \,,
	\end{aligned}
\end{align}
\begin{align}\label{eq:g1sol}
	\begin{aligned}
		g_{1} (r) = & \frac{2 \alpha _1 r^2}{3 L^4}-\frac{\left(65 \alpha _1+54 \alpha _2\right) q^2}{3 L_0^2 r^4}+\frac{2 \alpha _1 m^2}{r^6} - \frac{4 \left(4 \alpha _1+3 \alpha _2\right) q^2}{r^6} \\
		&  + \frac{\left(31 \alpha _1+30 \alpha _2\right) m q^2}{3 r^8} -\frac{\left(191 \alpha _1+72 \left(3 \alpha _2+2 \alpha _3+\alpha_4\right)\right) q^4}{96 r^{10}} \,,
	\end{aligned}
\end{align}
\begin{align}\label{eq:phi1sol}
	\Phi_1(r) = \frac{\left(-13 \alpha_1 + 24 \left(2 \alpha _2+6 \alpha _3+3 \alpha _4\right) \right) q^3}{8 \sqrt{3} r^8}
	-\frac{4 \sqrt{3} \alpha _2 q \left(L_0^2 m-r^4\right)}{L_0^2 r^6} \,.
\end{align}
To fix the integrating constants coming from the integration of the corrected equation of motion is done in such a manner that the corrected metric solution satisfies the asymptotic AdS solution as exemplified in \eqref{eq:vacmet2}, and we held one of the integrating constants to zero, demanding a shift in $m$ parameter.
A fixed charge configuration determines an integrating constant from Maxwell's equation. Where the electric charge is defined as a conserved Noether's charge given by
\begin{equation}\label{eq:Qdef}
	Q = \frac{1}{16 \pi G_N } \int_{S_3} \star \mathcal{F} \,,
\end{equation}
where $\mathcal{F}$ is the effective field strength
\begin{align}
	\mathcal{F}_{\mu \nu } = F_{\mu \nu } -4 \alpha' \Big(\alpha _2 F^{\alpha \beta } R_{\mu \nu \alpha \beta } +2 \alpha _3 F_{\alpha \beta } F^{\alpha \beta } F_{\mu \nu } +2 \alpha _4 F_{\alpha \beta } F_{\mu }{}^{\alpha } F_{\nu }{}^{\beta } \Big) \,.
\end{align}

\subsection*{Thermodynamic quantities with corrected geometry}
In this subsection, we compute the main thermodynamic quantities to describe the black hole as the Hawking temperature, Wald entropy and the free energy of the black hole. For a Euclideanized black hole, the periodicity $\beta$ of Euclidean time $\tau$ or the black hole temperature is given by
\begin{equation}\label{eq:hawkingtempdef}
	T = \frac{1}{\beta} = \frac{\kappa}{2 \pi} = \frac{1}{4\pi} \sqrt{g'(r) f'(r)} \bigg|_{r_+}.
\end{equation}
After some simplification, we find that the corrected Hawking temperature is given by
\begin{align}\label{eq:tempHD}
	\begin{aligned}
		T =  & \frac{r_+}{\pi  L_0^2} +\frac{1}{2 \pi  r_+} - \frac{q^2}{8 \pi  r_+^5}+\alpha ' \bigg[ \alpha _1 \bigg(\frac{q^2 \left(5 L_0^2-4 r_+^2\right)}{3 \pi  L_0^2 r_+^7} \\
		& -\frac{2 \left(3 L_0^4+6 L_0^2 r_+^2+2 r_+^4\right)}{3 \pi  L_0^4 r_+^3} 
		-\frac{9 q^4}{16 \pi r_+^{11}}\bigg) \\
		&  + \alpha _2 \left(\frac{q^4}{4 \pi  r_+^{11}}-\frac{q^2 \left(L_0^2+2 r_+^2\right)}{\pi  L_0^2 r_+^7}\right) + \frac{3 \left(2 \alpha _3+\alpha _4\right) q^4}{2 \pi r_+^{11}} \bigg] \,.
	\end{aligned}
\end{align}\\
For the generic 4-derivative theory shown in \eqref{eq:actionHD}, the Wald entropy tensor will take the form,
\begin{align}\label{eq:WentdenHD}
	\frac{\partial {\cal L}}{\partial R^{\alpha\beta\gamma\delta}} = \frac{1}{16\pi G_N}\bigg(\frac{1}{2}( g_{\alpha\gamma}g_{\beta\delta}- g_{\alpha\delta}g_{\beta\gamma}) 
	+ \alpha' \left( 2\alpha_1 R_{\alpha\beta\gamma\delta} +  \alpha_2 F_{\alpha\beta}F_{\gamma\delta} \right) \bigg) \,,
\end{align}
and using an appropriate definition of binormal tensor by following \cite{Dutta:2006vs}, the Wald entropy for the black hole is
\begin{align}\label{eq:entHD}
	S = \mathcal{S}_W = \frac{\pi ^2 r_+^3}{2 G_N} + \alpha ' \bigg[ \alpha _1 \bigg(\frac{2 \pi ^2 r_+ \left(3 L_0^2+2 r_+^2\right)}{G_N L_0^2} -\frac{7 \pi ^2 q^2}{2 G_N r_+^3}\bigg) -\frac{3 \pi ^2 \alpha _2 q^2}{G_N r_+^3} \bigg] \,.
\end{align}
Alternatively, the corrected entropy can be computed using Euclidean computation. Similarly, the ADM mass of the black hole can also be obtained by either computing the asymptotic stress tensor \cite{Arnowitt:1959ah, Balasubramanian:1999re} or on-shell Euclidean action \cite{Dutta:2006vs}. The result is given by
\begin{align}\label{eq:massHD}
	\begin{aligned}
		M & = \frac{3 \pi^2}{8 \pi G_N} \bigg[r_+^2 + \frac{r_+^4}{L_0^2} + \frac{q^2}{4 r_+^2}  + \frac{\alpha '}{L_0^2} \bigg( \alpha _1 \bigg(\frac{2 \left(3 L_0^4+12 L_0^2 r_+^2+10 r_+^4\right)}{3 L_0^2}+\frac{23 L_0^2 q^4}{32 r_+^8} \\
		& -\frac{14 q^2 \left(L_0^2+2 r_+^2\right)}{3 r_+^4} \bigg)+\alpha _2 \left(\frac{L_0^2 q^4}{4 r_+^8}-\frac{2 q^2 \left(L_0^2+4r_+^2\right)}{r_+^4}\right) -\frac{3 \left(2 \alpha _3+\alpha _4\right) L_0^2 q^4}{4 r_+^8} \bigg)\bigg] \,.
	\end{aligned}
\end{align}
And finally, we compute the free energy of the black hole from the on-shall action computation via holographic renormalisation or counterterm method:
\begin{equation}\label{eq:freeEdef}
	\beta F = S^{OS}_{tot} = S_{\text{bulk}} + S_{\text{bdy}} - S_{\text{ct}} \,.
\end{equation}
Thus the renormalised free energy of the black hole is 
\begin{align}\label{eq:freeEHD}
	\begin{aligned}
		F = & \frac{\pi  r_+^2 \left(L_0^2-r_+^2\right)}{8 L_0^2}+ \frac{\pi  \left(3 L_0^2 r_+^2+5 q^2\right)}{32 r_+^2} + \alpha ' \bigg[\alpha _1 \bigg( \frac{29 \pi  L_0^2 q^4}{256 r_+^8}+\frac{\pi  q^2 \left(14 r_+^2-L_0^2\right)}{12 r_+^4} \\
		& -\frac{\pi  \left(81 L_0^4+144 L_0^2 r_+^2+40 r_+^4\right)}{48 L_0^2} \bigg) +\alpha _2 \bigg( \frac{\pi  q^2 \left(5 L_0^2+4 r_+^2\right)}{4 r_+^4}-\frac{13 \pi  L_0^2 q^4}{32 r_+^8}  \bigg)\\
		& \qquad -\frac{33 \pi  \left(2 \alpha _3+\alpha _4\right) L_0^2 q^4}{32 r_+^8}\bigg] \,.
	\end{aligned}
\end{align}
Since we are working in canonical ensemble \emph{i.e.} fix charge configuration, the free energy upto $\mathcal{O} (\alpha ')$ satisfies the relations 
\begin{equation}
	F + T S = M = \frac{\partial S^{OS}_{{tot}}}{\partial \beta}, \qquad  \mathcal{S}_W  = \beta \frac{\partial S^{OS}_{{tot}}}{\partial \beta} - S^{OS}_{{tot}} \,.
\end{equation}

\subsection*{Holographic first law and the chemical potential}

Earlier, we examined in Section \ref{sec:firstlaw} how the first law of thermodynamics for charged black holes in the presence of higher derivative terms is modified. Their first law is illustrated as
\begin{align}
	d M = \frac{T }{4 G_N} dA + \Phi dQ + \frac{\Theta}{8 \pi G_N} d\Lambda_0  - \left(M - {\Phi Q} \right) \frac{dG_N}{G_N} + \frac{\mathcal{A}_{\alpha '}}{G_N} d \alpha ' \,,
\end{align}
where
\begin{align}\label{eq:thetaHD}
	\Theta = -\frac{1}{2} \pi ^2 r_+^4 + \alpha ' \bigg(\alpha _1 \left(\frac{4 \pi ^2 q^2}{r_+^2}-\frac{4 \pi ^2 r_+^2 \left(L_0^2+r_+^2\right)}{3 L_0^2}\right)
	+\frac{4\pi ^2 \alpha _2 q^2}{r_+^2}\bigg) \,, 
\end{align}
and
\begin{align}
	\begin{aligned}
		\mathcal{A}_{\alpha '} = \alpha _1 \bigg(\frac{\pi  q^2 \left(3 L_0^2+2 r_+^2\right)}{4 L_0^2 r_+^4}-\frac{\pi  \left(9 L_0^4+20 L_0^2 r_+^2+6 r_+^4\right)}{4 L_0^4} -\frac{43 \pi  q^4}{256 r_+^8}\bigg) \\
		+\alpha _2 \left(\frac{3 \pi  q^2}{4 r_+^4}-\frac{9 \pi  q^4}{32 r_+^8}\right) -\frac{9 \pi  \left(2 \alpha _3+\alpha _4\right) q^4}{32 r_+^8} \,.
	\end{aligned}
\end{align}
Thereafter, we want to recast the first law in terms of the boundary theory variable $\mathtt{c}_{\pm}$, as we discussed beforehand. Where $\mathtt{c}_{\pm}$ is constructed out of anomaly coefficient $\mathtt{c}$ and $\mathtt{a}$ in boundary CFT, which emerges from the presence of higher derivative terms. Following \cite{Banerjee:2009fm, Henningson:1998gx}, one can compute these anomaly coefficients:
\begin{align}\label{eq:candadefHD}
	\begin{aligned}
		\mathtt{c} & = \frac{L_0^3}{128 \pi G_N} \left( 1 + 3 \alpha_1 \frac{\alpha'}{L_0^2}\right) \\
		\mathtt{a} & = \frac{L_0^3}{128 \pi G_N} \left( 1 - 5 \alpha_1 \frac{\alpha'}{L_0^2}\right) \,.
	\end{aligned}
\end{align}
Thus the extended first law of thermodynamics in terms of $\mathtt{c_{\pm}}$ takes the form as 
\begin{align} \label{eq:bdyflHD}
	d E = T dS + {\tilde{\Phi}} d{\tilde{Q}} - p d\mathcal{V} + \mu_+ d\mathtt{c}_+ + \mu_- d\mathtt{c}_- \,,
\end{align}
where $p = M /3 \mathcal{V}$ represent the field theory pressure with boundary volume $\mathcal{V}$ and $\mu_{\pm} $ is given by
\begin{align}\label{eq:mupexpHD}
	\begin{aligned}
		\mu_+ & = \frac{8 \pi ^2 q^2}{L_0^3 r_+^2}+\frac{16 \pi ^2 r_+^2 \left(L_0^2-r_+^2\right)}{L_0^5} +  \frac{\alpha '}{L_0^2} \bigg[ \alpha _2 \bigg(\frac{8 \pi ^2 q^4}{L_0 r_+^8} -\frac{32 \pi ^2 q^2 \left(L_0^2-4 r_+^2\right)}{L_0^3 r_+^4}\bigg) \\
		& +\alpha _1 \bigg(-\frac{64 \pi ^2 q^2 \left(5 L_0^2-4 r_+^2\right)}{3 L_0^3 r_+^4} \\
		& +\frac{128 \pi ^2 \left(3 L_0^4+6 L_0^2 r_+^2+2 r_+^4\right)}{3 L_0^5}+\frac{23 \pi ^2 q^4}{L_0 r_+^8}\bigg) \qquad -\frac{24 \pi ^2 \left(2 \alpha _3+\alpha _4\right) q^4}{L_0 r_+^8} \bigg] \,,
	\end{aligned}
\end{align}
and
\begin{align}\label{eq:mudexpHD}
	\begin{aligned}
		\mu_- & = \frac{6 \pi ^2 q^2 \left(4 L_0^2+3 r_+^2\right)}{L_0^3 r_+^4}-\frac{4 \pi ^2 \left(18 L_0^4+39 L_0^2 r_+^2+13 r_+^4\right)}{L_0^5} \\
		& -\frac{43 \pi ^2 q^4}{8 L_0 r_+^8} + \frac{1}{\alpha _1} \bigg( \alpha _2 \left(\frac{24 \pi ^2 q^2}{L_0 r_+^4}-\frac{9 \pi ^2 q^4}{L_0 r_+^8}\right) -\frac{9 \pi ^2 \left(2 \alpha _3+\alpha _4\right) q^4}{L_0 r_+^8} \bigg) \,,
	\end{aligned}
\end{align}
and they satisfy the Euler relation of the boundary theory as presented in \eqref{eq:eulerHD}.


\section{Critical behaviour and Phase structure of bulk and boundary thermodynamics}\label{sec:phasestr}

By treating the cosmological constant as a thermodynamic pressure and its conjugate quantity as the volume, the phase structure was significantly enhanced, leading to an analogy between the liquid-gas and black hole systems. Considering Newton's constant as a thermodynamic parameter opens up an entirely novel perspective on the phase structure of bulk thermodynamics, which facilitates the engagement towards the study of critical phenomena and the phase space description of the boundary CFT. In this section, we discuss the phase structure of the charged black hole in $\text{AdS}_5$ with higher derivative correction and investigate the criticality behaviour of bulk theory and boundary theory.

\subsection{Gauss-Bonnet AdS Charge Black hole}

\subsubsection{Critical point analysis of $\text{AdS}_5$ Gauss-Bonnet black hole}
In canonical ensemble \textit{i.e.} fixed charge Q configuration,  one can interpret the equation \eqref{eq:Pdef} with \eqref{eq:tempGB} as the \emph{equation of state} as $P = P(r_+, T)$ for Gauss-Bonnet AdS black hole. Given the equation of state, one can smoothly calculate the critical point of the system. In the following equation,  we can compute the critical points.
\begin{equation}\label{eq:cripointdef}
	\frac{\partial T}{ \partial r_+} = 0 \,, \qquad \frac{\partial^2 T}{ \partial r_+^2} = 0 \,,
\end{equation}
solving the overhead equations exactly is challenging. However, we are interested in the solution's approximate behaviour to understand the universal behaviour of critical points and its dependence on the parameter $\alpha '$. Hence, we solve these equations perturbatively up to $\mathcal{O}(\alpha')$, we get 
\begin{align}\label{eq:cripointGB}
	\begin{aligned}
		r_{+(cri)} & = 15^{1/4} \sqrt{q} \left( 1 + \alpha ' \frac{8 \sqrt{3}}{5^{3/2} q} \right) \,, \\
		L_{(cri)} & = 3^{1/4} 5^{1/4} \sqrt{q} \left( 1 + \alpha ' \frac{16 \sqrt{3}}{5^{3/2} q} \right) \,.
	\end{aligned}
\end{align}
Setting the value of $r_{+(cri)}$ and $L_{(cri)}$ in the eq. \eqref{eq:tempGB} and eq. \eqref{eq:Pdef}, we can find the critical value of the bulk temperature and pressure. Thus the critical value of temperature and pressure is
\begin{align}\label{eq:tpcriGB}
	T_{cri} & = \frac{4}{5 \times 15^{1/4} \pi \sqrt{q}} \left( 1 - \alpha ' \frac{4 \sqrt{3}}{\sqrt{5} q}\right) \,, \nonumber \\
	P_{cri} & = \frac{1}{4 \sqrt{15} \pi G_N q} \left( 1 - \alpha ' \frac{278}{15 \sqrt{15} q} \right) \,,
\end{align}
and in the end, with the support of eq. \eqref{eq:candadefGB}, the critical value of central charges upto $\mathcal{O}(\alpha ')$ is given by  
\begin{align}
	\begin{aligned}
		\mathtt{c}_{cri} & = \frac{9 \times 3^{1/4} 5^{3/4} q^{3/2}}{128 \pi G_N } \left( 1 + \alpha ' \frac{397}{15\sqrt{15}q} \right) \,, \\
		\mathtt{a}_{cri} & = \frac{9 \times 3^{1/4} 5^{3/4} q^{3/2}}{128 \pi G_N } \left( 1 + \alpha ' \frac{119}{5\sqrt{15}q} \right) \,.
	\end{aligned}
\end{align}

\subsection*{Behavior near critical point}

In this subsection, we will compute the critical exponent for the charged Gauss-Bonnet black hole, which stands for the phase transition's universal property. In general, around the critical point, a VdW-like phase transition is characterized by the four critical exponents $\alpha, $ $ \beta, $ $ \gamma, $ and $\delta$, which are defined as
\begin{equation}
	\begin{aligned}
		& \text{Specific heat : }  C_v = \left. T \frac{\partial S}{\partial T} \right|_v \propto \left(-\frac{T-T_{cri}}{T_{cri}}\right)^{-\alpha} \,, \\[3pt]
		& \text{Order parameter : }  \eta=\frac{v_s-v_l}{v_{cri}} \propto\left(-\frac{T-T_{cri}}{T_{cri}}\right)^\beta \,, \\[3pt]
		& \text{Isothermal compressability:}  \kappa_T = -\left.\frac{1}{v} \frac{\partial v}{\partial P}\right|_T \propto\left(-\frac{T-T_{cri}}{T_{cri}}\right)^{-\gamma} \,, \\[3pt]
		& \text{Equation of state : }  P-P_{cri} \propto\left(v-v_{cri}\right)^\delta,
	\end{aligned}
\end{equation}
where $v$ be the specific volume.\footnote{We can identify the specific volume $v$ with the horizon radius of the black hole as $$ v = \frac{4 l_p^{1-d}}{d-1} r_+ $$.} Computation of these exponents is given in the appendices \ref{app:criexpo}. Our result for charged GB-AdS black hole is given by
\begin{equation}
	\alpha = 0 \,, \qquad \beta = \frac{1}{2}\,, \qquad \gamma = 1\,, \qquad \delta = 3 \,.
\end{equation}
Apparently, the critical exponents of the five-dimensional spherical charged GB-AdS black holes coincide with the computation of critical exponents from the mean-field theory for the Van der Waals liquid-gas system. We can write the thermodynamics quantities in terms of boundary variables by implementing the holographic dictionary. We obtained an equivalent result for the critical exponent for boundary field theory dual to charged GB-AdS black holes.

\subsection*{Phase transition}

\begin{figure}[h!]%
	\centering
	\subfloat[\centering]{{\includegraphics[width=7.0cm]{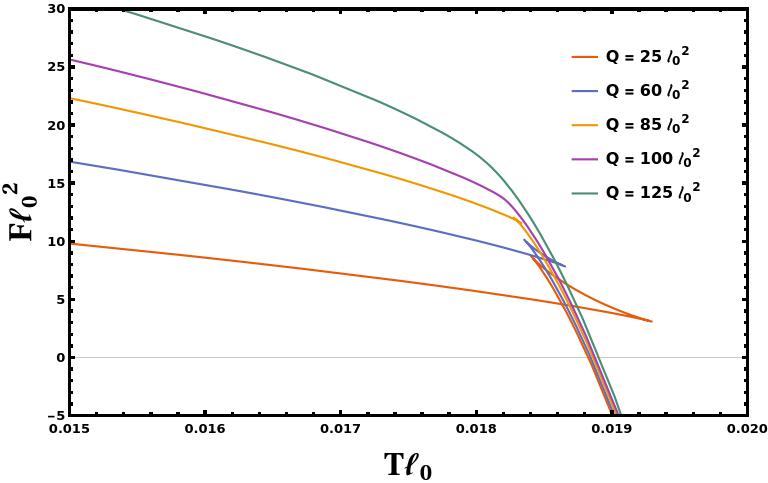} }}%
	\quad
	\subfloat[\centering]{{\includegraphics[width=7.0cm]{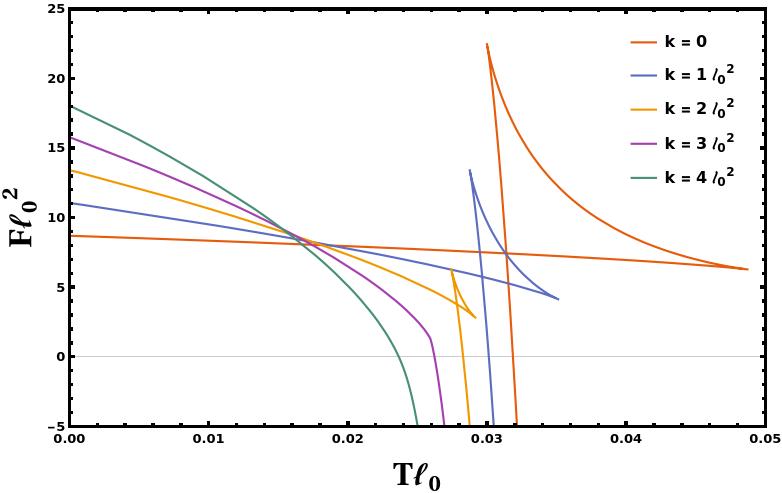} }}%
	\caption{Gibbs free energy $F$ vs. temperature $T$ diagram. Gibbs free energy $F \ell_0^2 $ is displayed as a function of temperature $T \ell_0$ in a different configuration of electric charge and different value of Gauss-Bonnet parameter. \textbf{(a)} We vary the total charge $Q$ of the black hole keeping other parameters to a fixed value as $ G_N = 1, L_0 = 25 \ell_0, \alpha' = 5 \ell_0^2 $. \textbf{(b)} We vary the Gauss-Bonnet parameter $\alpha ' \in (0, 1, 2, 3, 4)\ell_0^2$ holding other parameters to $G_N =1, L_0 = 25 \ell_0, Q = 20 \ell_0^2 $ .}%
	\label{fig:FvsTGB}%
\end{figure}

We mainly work in canonical ensembles with fixed $(Q)$ configurations. As we can express the free energy as a function of $T$, $P$ or $L_0$, $Q$, $G_N$ and higher derivative coupling  parameters.
\begin{align}\label{eq:freeEGB}
	F = \frac{\pi }{8 G_N L_0^2 r_+^2 \left(r_+^2 + 4 \alpha '\right)} \Big(q^2 L_0^2 \left(5 r_+^2 + 36 \alpha '\right) - 6 L_0^2 r_+^2 \alpha ' \left( r_+^2 - 4 \alpha '\right) + r_+^6 \left(L_0^2 - 36 \alpha '\right) - r_+^8\Big) \,.
\end{align}
In Fig \ref{fig:FvsTGB}, we plotted the Gibbs free energy $(F)$ given above with respect to the hawking temperature $(T)$ \eqref{eq:tempGB} of the corrected black hole for different value of $Q$ and $\alpha '$ while fixing the other parameters. There we introduce fiducial length $\ell_0$ such that all quantities are gauged in terms of $\ell_0$. We observe that the behaviour of the free energy is similar to Einstein's gravity as detailed studied in \cite{Chamblin:1999tk_charge_BH, Chamblin:1999tk_charge_BH_2, Kubiznak+Mann:2012wp_PVCriticality} and the Gauss bonnet parameter a significant role in the critical behaviour because it modifies the critical points. In Fig. \ref{fig:FvsTGB}, for $Q < Q_{crit}$ (Orange, blue), the free energy displays a “swallowtail” behaviour and a first-order phase transition emerges between two thermodynamically stable branches. The “horizontal” branch has low entropy, corresponding to the small black hole, while the “vertical” branch has high entropy reaching the massive black hole. Again the swallowtail-like behaviour or the first-order phase transitions occur for different Gauss-Bonnet parameters, and it is observed that swallowtail behaviour disappears for a large value of $\alpha'$.

\begin{figure}[h!]%
	\centering
	\subfloat[\centering]{{\includegraphics[width=7.0cm]{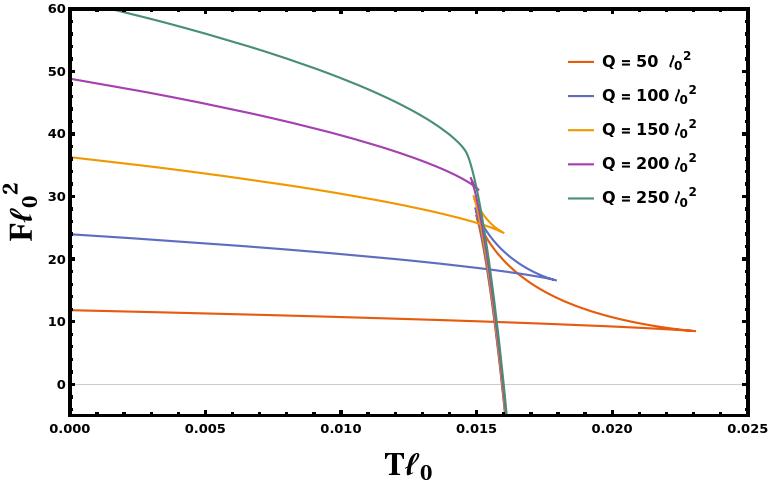} }}%
	\quad
	\subfloat[\centering]{{\includegraphics[width=7.0cm]{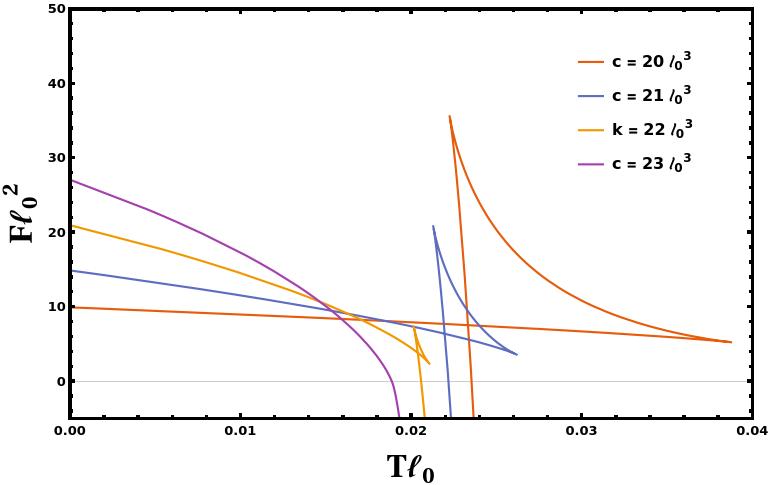} }}%
	\caption{Gibbs free energy $F$ vs. temperature $T$ diagram. Gibbs free energy $F \ell_0^2 $ is displayed as a function of temperature $T \ell_0$ in terms of boundary variable. \emph{Left :} We vary the total charge $ Q \in (50, 100, 150, 200, 250) \ell^2_0$ of the black hole keeping other parameters to a fixed value as $ L_0 = 30 \ell_0, \mathtt{c}_+ = 20 \ell_0^3, \mathtt{a}_- = 20 \ell_0^3 $. \emph{Right :} We vary the $\mathtt{c}_+ \in (20,21,22,23) \ell_0^3 $ holding other parameters to $ L_0 = 25 \ell_0, Q = 20 \ell_0^2 , \mathtt{c}_- = 20 \ell_0^3 $ .}%
	\label{fig:FvsTGBwithca}%
\end{figure}

Next, we rewrite the free energy $F$ in terms of boundary variable as $ F(T, L_0, Q, \mathtt{c}_+, \mathtt{c}_- ) $ and in Fig. \ref{fig:FvsTGBwithca} we study the critical behaviour of free energy w.r.t the boundary observable.

Likewise the bulk phase structure, we observed the swallowtail behaviour in the phase diagram of Free energy for various configurations of electric charge of boundary CFT while preserving the form of the central charge, and in fixed charge ensemble, the variation of $\mathtt{c_+}$ lead to swallowtail behaviour when $\mathtt{c_+} \sim \mathtt{c_-}$ and its start disappearing for the considerable contrast between the values of $\mathtt{c_+} $ and $\mathtt{c_-}$.

\begin{figure}[h!]%
	\centering
	\subfloat
	{{\includegraphics[width=7.0cm]{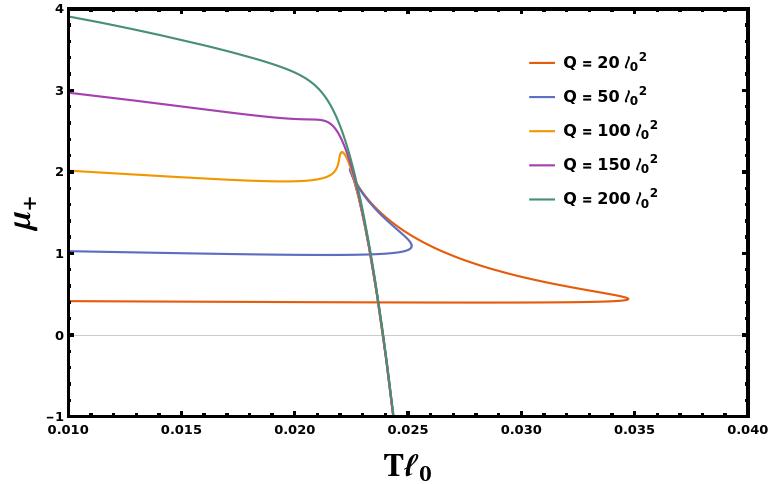} }}
	\quad
	\subfloat
	{{\includegraphics[width=7.0cm]{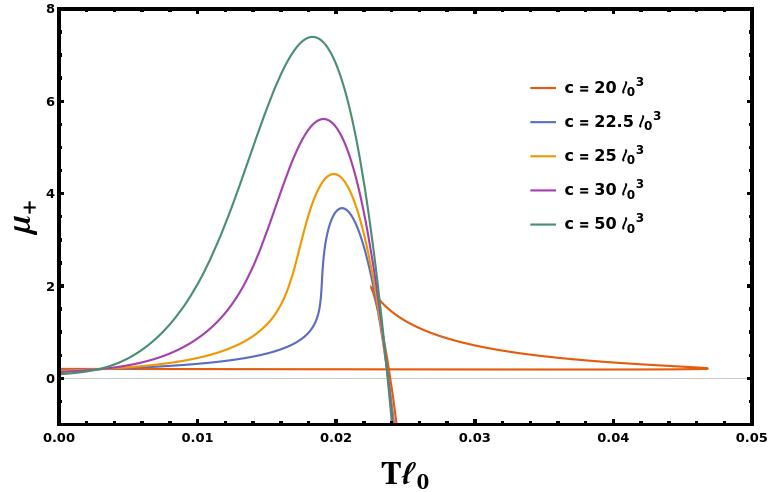} }}%
	\caption{ Chemical potential $\mu_+$ vs T diagram. The chemical potential conjugate to $\mathtt{c}_+$ is plotted against the temperature of the boundary theory in various configurations. \textbf{(a)} we vary the total electric charge $Q \in (20, 50, 100, 150, 200) \ell_0^2$ with a fixed value of $L_0 = 20 \ell_0^2 $, $\mathtt{c_+} = 10 \ell_0^2 $, $\mathtt{c_-} = 10 \ell_0^2 $. \textbf{(b)} we vary the $\mathtt{c}_+$ while keeping the fixed value of $L_0 = 20 \ell_0^2 $, $Q = 20 \ell_0^2 $, $\mathtt{c_-} = 20 \ell_0^2 $ } %
	\label{fig:muvsTwithc}%
\end{figure}

We investigate the phase behaviour of the chemical potential conjugate to the central charge of the boundary theory in Figure \ref{fig:muvsTwithc}. According to the phase diagram of free energy expressed in terms of boundary variables, we discovered that $\mu_+$ exhibits a swallowtail behaviour. Additionally, $\mu_+$ exhibits behaviour resembling the chemical potential $\mathcal{A}_{\alpha '}$ emerges in the presence of higher derivatives.

\subsection{Charged AdS Black hole in generic 4-derivative gravity}

Here we present the qualitative behaviour of the critical point of charged AdS black hole in the generic 4-derivative theory. One can find the critical points perturbatively using the equation given in \eqref{eq:cripointdef}. The critical point in the 4-derivative theory are shown below
\begin{align}\label{eq:cripointHD}
	r_{+(cri)} & = c \sqrt{q} \left( 1 + \alpha ' \frac{\left(398 \alpha _1+672 \alpha _2-528 \left(2 \alpha _3+\alpha_4\right)\right) }{75 \sqrt{15} q}\right) \,, \\
	L_{(cri)} & = c \sqrt{3 q} \left( 1 +  \alpha ' \frac{\left(383 \alpha _1+192 \alpha _2-88 \left(2 \alpha _3+\alpha _4\right)\right)}{25 \sqrt{15} q} \right) \,.
\end{align}
where $c=15^{1/4}/\sqrt{2}$. Putting the value of $r_{+(cri)}$ and $L_{(cri)}$ in the eq. \eqref{eq:tempHD} and eq. \eqref{eq:Pdef}, we can find the critical value of the bulk temperature and pressure. Thus the critical value of temperature and pressure is
\begin{align}\label{eq:tpcriHD}
	T_{cri} & = \frac{4}{5 \pi c \sqrt{q}} \left( 1 - \alpha ' \frac{\left(934 \alpha _1+336 \alpha _2-144 \left(2 \alpha _3+\alpha
		_4\right)\right)}{45 \sqrt{15} q} \right) \,, \\
	P_{cri} & = \frac{1}{4 \pi G_N c^2 q} \left( 1 - \frac{\left(6794 \alpha _1+3456 \alpha _2-1584 \left(2 \alpha _3+\alpha_4\right)\right) \alpha '}{225 \sqrt{15} q} \right) \,,
\end{align}
and in the end, with the support of eq. \eqref{eq:candadefHD}, the critical value of central charges upto $\mathcal{O}(\alpha ')$ is given by  
\begin{align}
	c_{cri} & = \frac{3\sqrt{3} c^3 q^{3/2}}{128 \pi G_N} \left(1 + \frac{ \alpha ' \left(1199 \alpha _1+576 \alpha _2-264 \left(2 \alpha _3+\alpha
		_4\right)\right) }{25 \sqrt{15} q} \right) \,, \\
	a_{cri} & = \frac{3\sqrt{3} c^3 q^{3/2}}{128 \pi G_N} \left( 1+ \frac{ \alpha ' \left(3197 \alpha _1+1728 \alpha _2-792 \left(2 \alpha _3+\alpha
		_4\right)\right) }{75 \sqrt{15} q}\right) \,.
\end{align}

\begin{figure}[h!]%
	\centering
	\subfloat[\centering]{{\includegraphics[width=7cm]{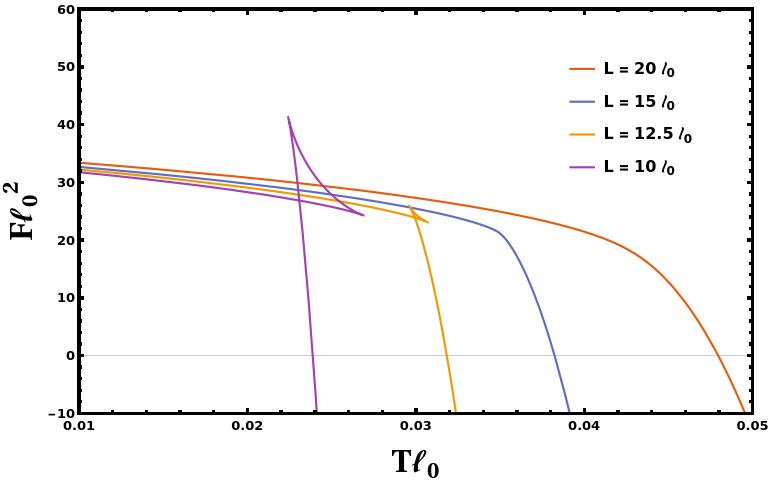} }}%
	\qquad
	\subfloat[\centering]{{\includegraphics[width=7cm]{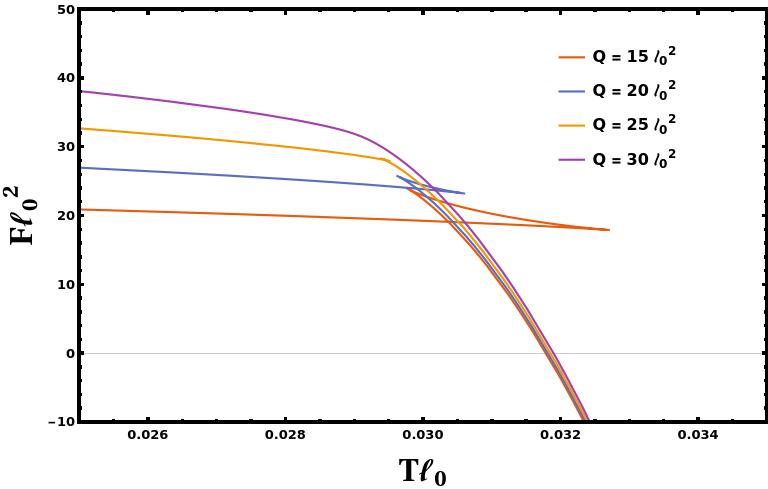} }}%
	\caption{Gibbs free energy $F$ vs. temperature $T$ diagram. Gibbs free energy $F \ell_0^2 $ is displayed as a function of temperature $T \ell_0$ in a different configuration of electric charge and AdS length. \textbf{(a)} We vary the total AdS curvature scale $L$ keeping other parameters to a fixed value as $ G_N = 1, Q = 20 \ell_0^2, \alpha' = 0.001 \ell_0^2 $. \textbf{(b)} We vary the total charge $Q$ of the black hole keeping other parameters to a fixed value as $ G_N = 1, L = 25 \ell_0, \alpha' = 0.001 \ell_0^2 $.}%
	\label{fig:FvsTHD}%
\end{figure}

We can express the free energy as a function of $T$, $P$ or $L$, $Q$, $G_N$ and higher derivative coupling  parameters. In Fig \ref{fig:FvsTHD}, we plotted the free energy $ (F) $ given in \eqref{eq:freeEHD} with respect to the hawking temperature $(T)$ \eqref{eq:tempHD} of the corrected black hole for different value of $P$ and $Q$ while fixing the other parameters. We observe that the behaviour of the free energy is the same as Einstein's gravity as detailed studied in \cite{Chamblin:1999tk_charge_BH, Chamblin:1999tk_charge_BH_2, Kubiznak+Mann:2012wp_PVCriticality} for the small coupling term of the higher derivative terms.

\vspace{1em}
\section{Conclusion and Discussion}\label{sec:conclusion}
\vspace{1em}
In this article, we consider the consequence of the higher derivative term (containing a higher derivative in the metric field as well as in the gauge field) on the charge black hole thermodynamics and uncover its consistency with the thermodynamics of the boundary field theory by implementing the AdS/CFT dictionary. These higher derivative terms naturally emerge in theories of quantum gravity, specifically in string theories, when we consider their low-energy limits. To study this further, we apply the effective field theory approach to the four derivative terms discussed in Section \ref{sec:HDexamp}.

Since bulk thermodynamics for black holes in gauss-bonnet gravity \cite{Cai:1998vy, Cai:2001dz, Cai:2013qga, Cvetic:2001bk, Nojiri:2001aj} and the 4-derivative effective action \cite{Cremonini:2019wdk, Mandal:2022ztj} discussed in \eqref{eq:actionHD} are well investigated, we discuss the extended structure of black hole thermodynamics and its holographic counterpart using variations of Newton's constant as well as variations of the cosmological constant. The emergence of the higher derivative terms in bulk comes with an additional parameter $\alpha'$. The implication of the well-established AdS/CFT dictionary leads the variation of $\Lambda$ to induce a variation in the 't Hooft coupling $\lambda$, apart from variations in colour $N$ and boundary volume $V$. Therefore, to disentangle the $\lambda$ variation from that of $N$ and $V$, we allow the parameter $\alpha'$ to vary in bulk along with $L$ and $G_N$ as bookkeeping devices. As a result, we will be able to demonstrate that boundary and bulk thermodynamics are indeed equivalent.

We will proceed with the general four-derivative theory of gravity coupled with the U(1) gauge field in the bulk black holes. In the presence of such terms, we include the variation of $\alpha'$ in the bulk first law and show that the variations of $G_N$ and $\alpha'$ generate the variations of $c_+$ and $c_-$, where $c_\pm = (c \pm a)/2$, the bulk first law can be beautifully interpreted as the boundary first law which is written in terms of variations of $c_\pm$. As a result, the boundary theory is endowed with two chemical potentials $\mu_\pm$ (corresponding to $c_\pm$ respectively), and they satisfy the generalised Euler relation \eqref{eq:eulerHD} of the boundary theory. The existence of a gauge field modifies the general expression of the chemical potential $\mu_{\pm}$ as shown in \eqref{eq:mupmexp}. Furthermore, we concluded that the authors of \cite{Kumar:2022afq, Qu:2022nrt} did not consider the anomaly present in the boundary theory due to the presence of the higher derivative terms in the bulk theory, and as a result, they had not taken into account the other central charge contribution in the boundary theory's first law thermodynamics.

After establishing a one-to-one correspondence between bulk and boundary thermal parameter space, we study the phase behavior of the charged black hole in the higher derivative gravity theory. The phase structure of charged black holes in AdS with higher derivative terms is endowed with a different chemical potential; hence the dimension of the thermodynamic phase will increase. In this paper, we study thermodynamics perturbatively. However, in the first example, we discuss the Gauss-Bonnet AdS black holes, for which one can extract the complete solution. Therefore, without any perturbative analysis, we discuss the thermodynamic and phase behavior of a charged Gauss-Bonnet AdS black hole from both the bulk and boundary perspectives. We notice the swallowtail behavior in the Gauss-Bonnet AdS black hole, which is analogous to the Rissner-Nordstrom AdS black hole. In contrast, the swallowtail behavior will depend on both the charge of the black hole as well as on the Gauss-Bonnet parameter. The boundary CFT dual to Gauss-Bonnet black hole thermodynamics observed a similar phase structure. The analysis of the critical points generated by Equation \eqref{eq:cripointdef} is the crucial step in the study of phase structure. We found the critical points for charged Gauss-Bonnet black holes are the same as the critical point from the mean-field theory computation. One can see the dependence on the Gauss-Bonnet parameter on the critical points from \eqref{eq:cripointGB} and \eqref{eq:tpcriGB}. In the subsequent example, we study the thermodynamic phases of charged black holes in generic 4-derivative perturbatively.

It would also be interesting to find an adequate Van-der-Waals type description of higher derivative black holes and understand the effect of the central charges on the mean-field potential \cite{Dutta:2021whz}. Subsequently, it would be fascinating to investigate the validity of this formalism for rotating black holes or black holes with scalar charges in higher derivative theories \cite{Cano:2019ore,  Burger:2019wkq, Gao:2021ubl}, boosted black string \cite{Henriquez-Baez:2022bfi}, quantum black hole \cite{Pourhassan:2022opb, Pourhassan:2020yei}.

\vspace{1em}
\centerline{***************}
\vspace{1em}
\noindent
\textbf{Acknowledgement :} The authors would like to thank Suvankar Dutta and Arnab Rudra for their insightful comments on the first draft and numerous insightful discussions. GSP would like to thank the Department of Science \& Technology (DST) Fellowship for pursuing Ph.D. at IISERB. Finally, we owe the Indian people gratitude for their unwavering support of basic science research.

\appendix 
\section{Thermal quantities for Maxwell Gauss-Bonnet in term of $L$}\label{app:GBthermalqu}
Using the relation between $L$ and $L_0$, the Hawking temperature from \eqref{eq:tempGB} can be expressed as  
\begin{align}
	T = \frac{3 \pi ^2 r_+^4 \left(L^4+2 L^2 r_+^2-4 r_+^2 \alpha '\right)-G^2 L^4 Q^2}{6
		\pi ^3 L^4 r_+^3 \left(4 \alpha '+r_+^2\right)} \,,
\end{align}
and from \eqref{eq:massGB} the mass $M$ of black hole is
\begin{align}\label{eq:massGB2}
	M = \frac{3 \pi  \left(L^4 \left(2 \alpha '+r_+^2\right)+L^2 r_+^4-2 r_+^4 \alpha
		'\right)}{8 G L^4}+\frac{G Q^2}{8 \pi  r_+^2} \,.
\end{align}
The volume of the black hole as conjugate quantity correspond to $P$ defined in \eqref{eq:Pdef} and chemical potential conjugate to $\alpha'$ are given by
\begin{align}
	\left. \frac{\partial M}{\partial P} \right|_{r_+, Q, \alpha'} & = \frac{1}{2} \pi ^2 r_+^4-\frac{2 \pi ^2 r_+^4 \alpha '}{L^2}\,, \\
	\left. \frac{\partial M}{\partial \alpha'} \right|_{r_+, L, Q} & = -\frac{3 \pi ^2 \left(L^4 \left(3 r_+^2-4 \alpha'\right)+8 L^2 r_+^4-12 r_+^4 \alpha '+r_+^6\right)}{4 \pi  G L^4 \left(4 \alpha '+r_+^2\right)} \nonumber \\
	& + \frac{4 G^2 Q^2}{4 \pi  G r_+^2 \left(4
		\alpha '+r_+^2\right)}
\end{align}
Using the black hole mass \eqref{eq:massGB2}, we can generate the first law of thermodynamics and Smarr relation presented in \eqref{eq:firstlawHD} and \eqref{eq:smarrHD}, where $\Lambda_0$ is replaced with $\Lambda$, defined in term of the effective AdS length.

\section{Critical exponent Computation}\label{app:criexpo}

To find the critical exponent, we observed that $C_v = 0$ for the black holes and hence the first critical exponent is $ \alpha = 0 $. To discover the other exponent, we introduced the expansion parameter as $ t = \frac{T}{T_{cri}} - 1, \epsilon = \frac{v}{v_{cri}} - 1 $ and expand the equation of state near the critical point
\begin{equation}
	p = 1 + a_1 t + a_2 t \epsilon + a_3 \epsilon^3 + \mathcal{O}(t \epsilon^2, \epsilon^4)
\end{equation}
Using Maxwell's area law, during the phase transition
\begin{equation}
	\int_{\epsilon_l}^{\epsilon_s} \epsilon \frac{d p}{d\epsilon} d\epsilon = 0 \implies \epsilon \propto \sqrt{-t} \,.
\end{equation}
Therefore, we have
\begin{equation}
	\eta=\frac{v_s-v_l}{v_{cri}} \propto \sqrt{-t} \implies \beta = 1/2 
\end{equation}
The isothermal compressibility gives us the third critical exponent as 
\begin{equation}
	\kappa_T = - \frac{1}{v} \left.  \frac{ \partial v}{\partial P} \right| _{v_cri}  \propto - \left. \frac{1}{\frac{ \partial p}{\partial \epsilon}} \right|_{\epsilon = 0} =  \frac{1}{t}
\end{equation}
which indicates that the critical exponent $\gamma = 1$. Moreover, the shape of the critical isotherm t = 0
gives the fourth exponent as
\begin{equation}
	p - 1 \propto - \epsilon^3 \implies \delta = 3
\end{equation}

\printbibliography
\end{document}